\documentclass[lettersize,journal]{IEEEtran}
\usepackage{makecell}
\usepackage{amsmath,amsfonts}
\usepackage{dsfont}
\usepackage{algorithm}
\usepackage{algpseudocode}
\usepackage{hyperref}
\usepackage{cleveref}
\usepackage{array}
\usepackage{courier}
\hypersetup{hidelinks}
\usepackage{threeparttable} % for footnote in table
\usepackage{subfigure}
\usepackage{textcomp,soul}
\usepackage{stfloats}
\usepackage{url}
\usepackage{verbatim}
\usepackage{graphicx}
\usepackage{booktabs}
\usepackage{multirow}
\usepackage{cite}
\usepackage{amsthm}
\usepackage{bm}
\usepackage{amssymb}
\usepackage{xspace}

\theoremstyle{definition}

\usepackage[table]{xcolor}
\makeatletter
\pretocmd\@bibitem{\color{black}\csname keycolor#1\endcsname}{}{\fail}
\newcommand\citecolor[1]{\@namedef{keycolor#1}{\color{blue}}}
\makeatother 

\title{Diffusion Models for Intelligent Transportation Systems: A Survey} 

\author{Mingxing Peng, Kehua Chen, Xusen Guo, Qiming Zhang, Hui Zhong, Meixin Zhu*, and Hai Yang
\thanks{Manuscript received XX September, 2024.}
\thanks{Corresponding author is Meixin Zhu (E-mail: meixin@seu.edu.cn).}
\thanks{Mingxing Peng, Xusen Guo, Qiming Zhang, and Hui Zhong are with the Systems Hub, The Hong Kong University of Science and Technology (Guangzhou). Meixin Zhu is with the School of Transportation, Southeast University, Nanjing 211189, PR China. Kehua Chen is with Department of Civil and Environmental Engineering, University of Washington, Seattle, United States. Hai Yang is with the Department of Civil and Environmental Engineering, the Hong Kong University of Science and Technology, Clear Water Bay, Kowloon, Hong Kong, China.}}

\begin{document}
\maketitle
\begin{abstract}
Intelligent Transportation Systems (ITS) play a crucial role in enhancing traffic efficiency and safety. Recently, diffusion models have emerged as transformative tools for addressing the complex challenges faced within ITS. This paper presents a comprehensive survey of diffusion models in ITS, exploring both theoretical and practical dimensions. We begin by introducing the theoretical foundations of diffusion models and their key variants, such as conditional and latent diffusion models, emphasizing their capacity to model intricate, multi-modal traffic data and enable controllable generation. Next, we outline the primary challenges in ITS and the advantages diffusion models provide, facilitating a deeper understanding of the intersection between diffusion models and ITS. We then conduct a multi-perspective examination of current applications of diffusion models across ITS domains, including autonomous driving, traffic simulation, traffic forecasting, and traffic safety. Finally, we discuss state-of-the-art diffusion model techniques and highlight key research directions within ITS that merit further exploration. Through this structured overview, we aim to equip researchers with a comprehensive understanding of diffusion models in ITS, thereby fostering their future applications in the transportation domain.

\end{abstract}

\begin{IEEEkeywords}
Intelligent Transportation Systems, Diffusion Models, Autonomous Driving, Traffic Simulation, Traffic Forecasting, Traffic Safety.
\end{IEEEkeywords}

\begin{figure*}
\centering
\includegraphics[width=\linewidth, trim= 0 0 0 0, clip]{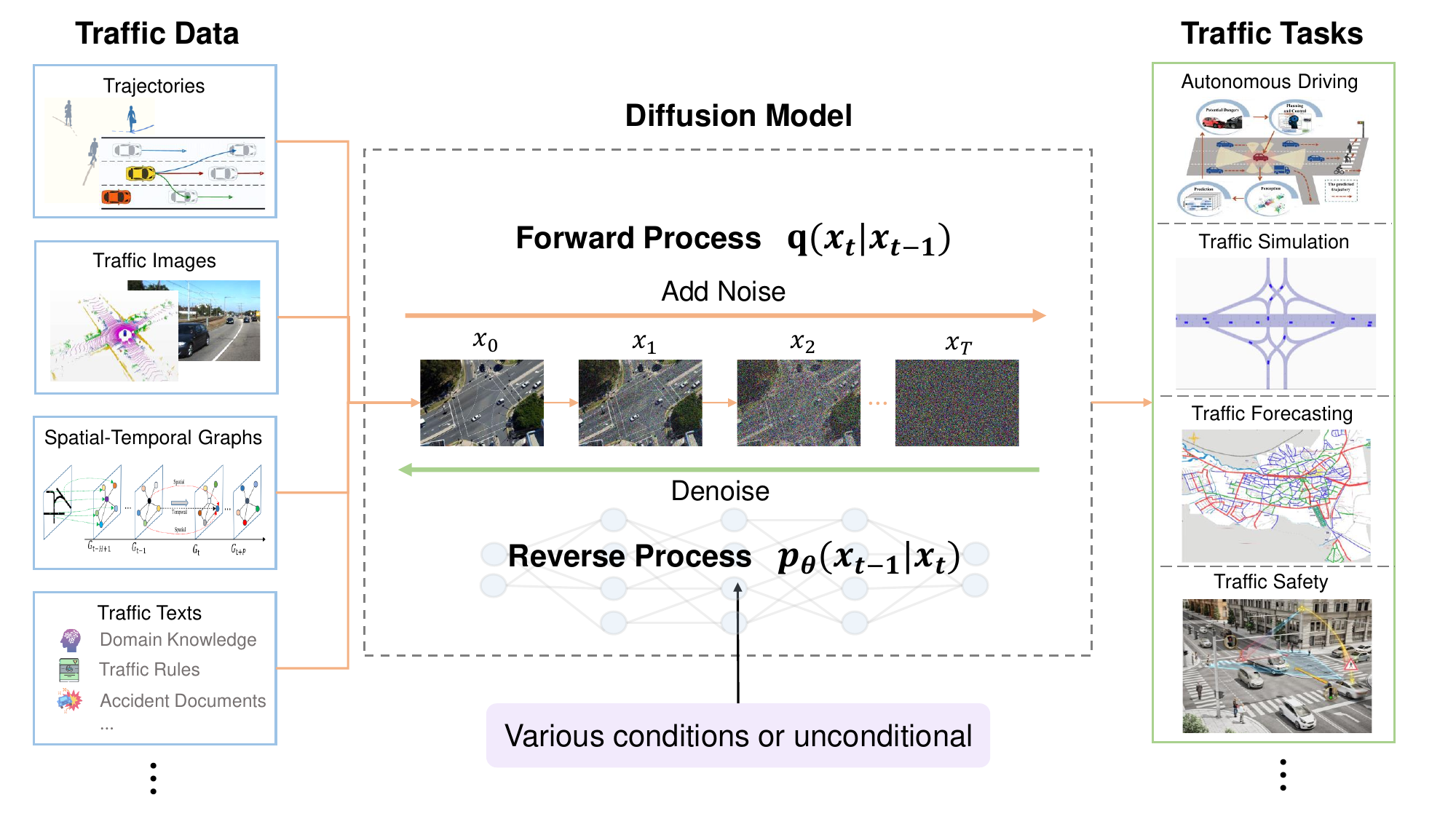}
\caption{Overview of applying diffusion models to traffic tasks using various traffic data types, including trajectories, traffic images, spatial-temporal graphs, and traffic-related texts.}
\label{fig:traffic_diffusion}
\end{figure*}

\section{Introduction}
\IEEEPARstart{A}{s} urbanization accelerates and populations grow, the demand for public transportation services increases alongside a steep rise in vehicle numbers. These trends have gradually revealed several issues in current transportation systems, such as traffic congestion and accidents. With advancements in computer technologies and transportation systems, many cities are increasingly focused on developing intelligent transportation systems (ITS) \cite{wootton1995intelligent}, which leverage cutting-edge technologies and extensive traffic data to enable efficient, high-quality, and safe traffic management. ITS encompasses several domains, including autonomous driving, which enhances traffic safety and efficiency; traffic simulation, which enables modeling, analysis, and testing of various strategies; traffic forecasting, which aims to reduce congestion and optimize services; and traffic safety, which seeks to minimize accidents and improve overall safety.

Traffic data are inherently heterogeneous and multi-modal, including vehicle and pedestrian trajectories, driving images or videos, spatial-temporal graphs derived from GPS positions, and textual data such as traffic rules and accident reports. These data often exhibit complex spatial-temporal dependencies and uncertainties. Additionally, the data may be noisy, incomplete, or difficult to obtain, with privacy concerns particularly affecting personal GPS data collection. Consequently, processing these multi-modal, complex, and often imperfect datasets presents a significant challenge for ITS.

In the past few decades, researchers have employed various approaches to address the challenges of ITS. For example, Recurrent Neural Networks (RNNs) are often used to model temporal relationships, while Convolutional neural networks (CNNs) are commonly utilized to capture spatial structure \cite{veres2019deep}. And graph-based approaches have demonstrated superior capabilities in extracting spatial correlations within traffic networks \cite{ye2020build, li2024survey}. However, these approaches often exhibit limitations when handling noisy or incomplete data. In contrast, generative models such as Generative Adversarial Networks (GANs) and Variational Autoencoders (VAEs) have proven effective for traffic data generation and imputation tasks \cite{lin2023generative, boquet2020variational}. However, GANs suffer from unstable training, and VAE has the limitation of low-quality output. As a powerful class of generative models, diffusion models offer advantages such as ease of training, enhanced generative performance, controllable generation, and multi-modal capabilities. To date, diffusion models have been applied across a wide range of vision tasks \cite{croitoru2023diffusion}, with promising applications such as Sora \cite{videoworldsimulators2024}. Inspired by these developments, an increasing number of researchers in the ITS domain have begun to adopt diffusion models to address various challenges in ITS. Therefore, originating in image processing and computer vision, diffusion models are now being applied across various traffic tasks, including autonomous driving, traffic simulation, traffic forecasting, and traffic safety. As illustrated in Fig.~\ref{fig:traffic_diffusion}, diffusion models are suitable for processing various traffic data and can address a wide range of traffic tasks based on task-specific conditions or unconditional methods.

There have been numerous surveys on intelligent transportation systems (ITS), focusing on deep learning techniques \cite{veres2019deep, khalil2024advanced}, along with focused surveys on graph-based learning methods \cite{ye2020build, li2024survey} and generative AI approaches \cite{yan2023survey} for ITS applications.
Meanwhile, several reviews have focused on diffusion models \cite{yang2023diffusion, jiang2024survey, yang2024survey} and their applications in areas such as computer vision \cite{croitoru2023diffusion} and medical imaging \cite{kazerouni2023diffusion}. However, there is currently no comprehensive review of diffusion models within the ITS domain. 

\begin{figure*}
\centering
\includegraphics[width=\linewidth, trim= 130 30 130 30, clip]{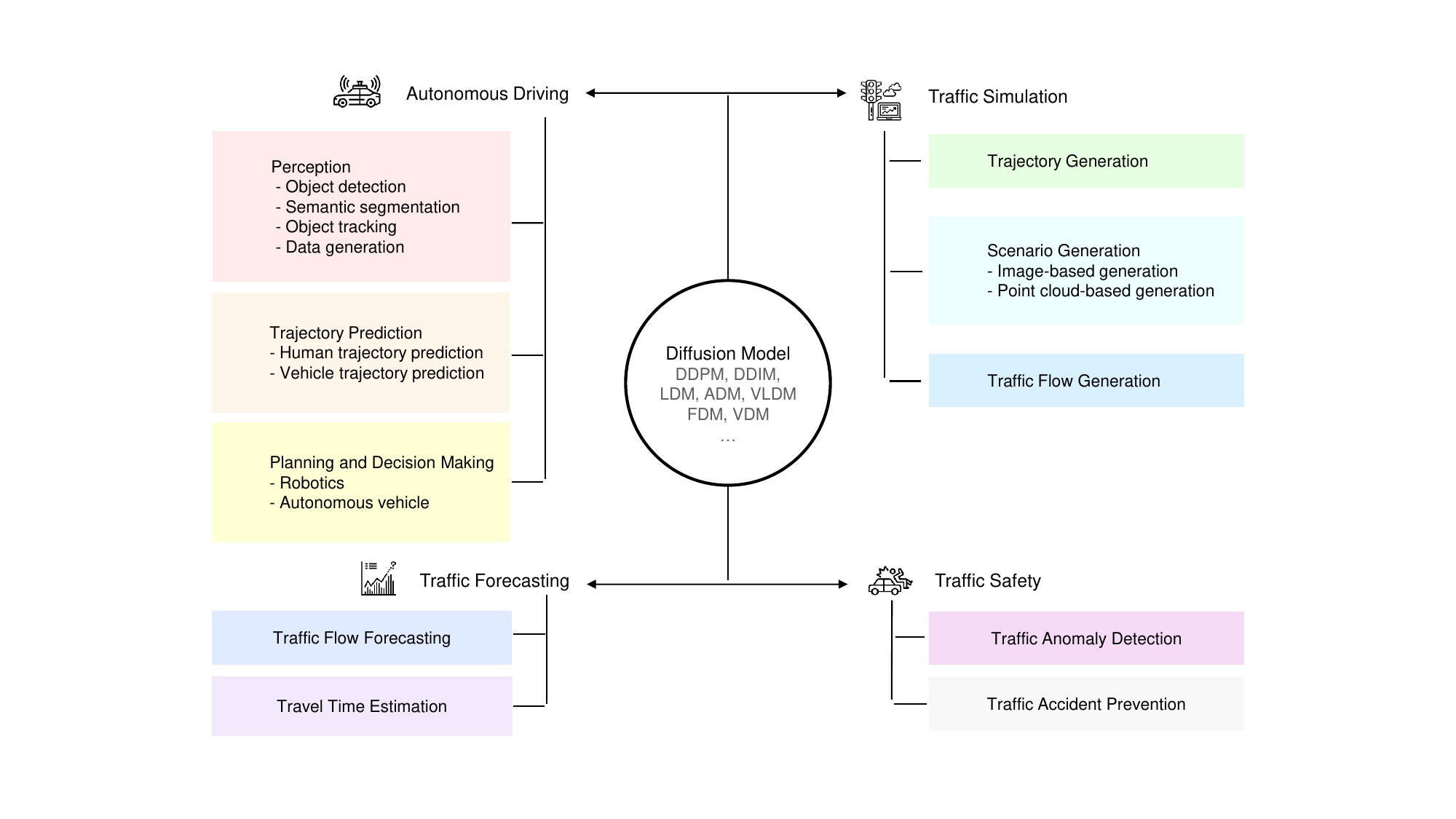}
\caption{Overview of the application of diffusion models in various domains of intelligent transportation systems.}
\label{fig:applications}
\end{figure*}

To address this gap, this paper presents a detailed literature review on diffusion models in ITS. First, we outline how diffusion models have emerged as powerful tools for various traffic tasks. Specifically, we introduce the theoretical foundations of diffusion models, along with conditional diffusion models and latent diffusion models, which extend their applicability to more specific tasks within ITS. Second, we examine the critical challenges in ITS and the corresponding advantages of diffusion models. Third, we investigate the applications of diffusion models in areas such as autonomous driving, traffic simulation, traffic forecasting, and traffic safety within ITS, as shown in Fig.~\ref{fig:applications}. In particular, we review these applications based on criteria such as task, denoising condition, or model architecture, as illustrated in Table.~\ref{tab:tab_taxonomy}. Finally, we provide an outlook on potential future directions for diffusion models in ITS. Our goal is to bridge the gap between the diffusion model and transportation research communities, fostering interdisciplinary collaboration and advancing the application of diffusion models in transportation.

In summary, the main contributions of this paper include: 
\begin{itemize}
    \item To the best of our knowledge, this is the first comprehensive literature review focused on the application of diffusion models in ITS.
    \item We systematically introduce how diffusion models have become powerful approaches for various traffic tasks by processing multi-modal and complex traffic data. Additionally, we explore the critical challenges in ITS and the corresponding advantages of diffusion models. This analysis offers readers more profound insights into the intersection of ITS and diffusion models.
    \item We present a comprehensive and up-to-date literature review of diffusion models in the ITS domain, focusing on applications in autonomous driving, traffic simulation, traffic forecasting, and traffic safety. By analyzing these applications through multiple perspectives, we aim to offer researchers from various ITS subfields a clear and efficient overview of the latest advancements in diffusion models.
    \item We discuss the cutting-edge techniques in diffusion models and highlight key research directions for diffusion models in ITS that are worthy of further exploration.
\end{itemize}

The remainder of the paper is organized as follows: Sec.~\ref{sec:theory} presents theoretical foundations of diffusion models and their key variants. Sec.~\ref{sec:challenge} outline the key challenges in ITS and the corresponding advantages of diffusion models. Sec.~\ref{sec:autonomous}- Sec.~\ref{sec:traffic_safety} explores the diverse applications of diffusion models within ITS, including autonomous driving, traffic simulation, traffic forecasting, and traffic safety. Sec.~\ref{sec:future} discusses several promising directions for future research. Finally, the conclusions are drawn in Sec.~\ref{sec:conclusion}.

\section{THEORY}
\label{sec:theory}
Diffusion models have emerged as transformative tools in the field of ITS. This section outlines how diffusion models have become powerful and flexible methods for addressing various traffic-related challenges. First, we explore the theoretical foundations of diffusion models, which lie in their ability to learn the underlying data distribution through a process of noise injection and subsequent denoising. This makes them highly effective for modeling complex traffic dynamics. Next, we introduce key variants of diffusion models, particularly conditional and latent diffusion models, which extend their applicability to more specific and challenging tasks within ITS. By incorporating domain-specific conditions and leveraging latent spaces, diffusion models can be applied to multi-modal traffic data, offering solutions to a wide range of traffic-related tasks.

\subsection{Foundations of Diffusion Models} 
Diffusion models are a powerful class of probabilistic generative models that gradually perturb data by adding Gaussian noise to data and then learn to reverse this process to generate new data. During training, the model learns to denoise the data at each step, effectively transforming random noise into coherent and realistic outputs.

This section provides an overview of three predominant formulations in diffusion models: Denoising Diffusion Probabilistic Models (DDPMs), which utilize discrete steps to add and remove noise incrementally; Noise Conditioned Score Networks (NCSNs), which estimate the gradient of the log-density of the data distribution to guide sample generation; and Stochastic Differential Equations (SDEs), which offer a continuous-time perspective that unifies and generalizes both DDPMs and NCSNs under a common mathematical framework. 

\vspace{5px}
\subsubsection{\textbf{Denoising Diffusion Probabilistic Models (DDPMs)}} \

DDPMs \cite{sohl2015deep, ho2020denoising} utilize two Markov chains: a forward (diffusion) process that gradually adds Gaussian noise to data, transforming it into pure noise over multiple steps, and a reverse (denoising) process, learning through neural networks—typically based on a U-Net architecture \cite{ronneberger2015u}—that progressively removes the noise to reconstruct the original data.

\textbf{Forward (Diffusion) Process.} The forward (diffusion) process incrementally corrupts the data by adding Gaussian noise in a series of $T$ steps. Given a data distribution $x_0 \sim q(x_0)$, the forward process starts with the original data $x_0$ and generates a sequence of latent variables $x_1, x_2, \dots, x_T$ through different diffusion steps. The process is defined by a Markov chain where each state $x_t$ depends only on the previous state $x_{t-1}$:

\begin{equation}
q({x}_t | {x}_{t-1}) = \mathcal{N}({x}_t; \sqrt{1 - \beta_t} {x}_{t-1}, \beta_t \mathbf{I}), \forall t \in \{1, \dots, T\}
\end{equation}
where $\beta_t \in (0, 1)$ is a hyperparameter representing the noise variance schedule that controls the amount of noise added at each step. $\mathbf{I}$ denotes the identity matrix, and $\mathcal{N}(x; \mu, \sigma)$ represents a normal distribution with mean $\mu$ and covariance $\sigma$.

The entire forward process can be expressed directly in terms of the original data ${x}_0$ using the reparameterization trick:

\begin{equation}
{x}_t = \sqrt{\bar{\alpha}_t} {x}_0 + \sqrt{1 - \bar{\alpha}_t} \epsilon, \quad \epsilon \sim \mathcal{N}(0, \mathbf{I})
\end{equation}
where $\alpha_t = 1 - \beta_t$ and $\bar{\alpha}_t = \prod_{s=1}^{t} \alpha_s$. 

\textbf{Reverse (Denoising) Process.} DDPMs aim to learn the reverse of this diffusion process, where the model starts with Gaussian noise and progressively removes the noise to generate new data. The reverse process is also modeled as a Markov chain, but it is parameterized by a neural network $p_\theta({x}_{t-1} | {x}_t)$ that generates $p_\theta({x}_{0})$ in a step-by-step manner:

\begin{equation}
p_\theta({x}_{t-1} | {x}_t) = \mathcal{N}({x}_{t-1}; \mathbf{\mu}_\theta({x}_t, t), \sigma^2 \mathbf{I})
\end{equation}

In the DDPM \cite{ho2020denoising}, the covariance $\sigma^2$ is fixed to a constant value, and the mean $\mathbf{\mu}_\theta({x}_t, t)$ is reformulated as:

\begin{equation}
\mathbf{\mu}_\theta({x}_t, t) = \frac{1}{\sqrt{\alpha_t}} \left({x}_t - \frac{\beta_t}{\sqrt{1 - \bar{\alpha}_t}} \epsilon_\theta({x}_t, t)\right)
\end{equation}
where $\epsilon_\theta({x}_t, t)$ represents the neural network's prediction of the noise component at step $t$.

The objective of training a DDPM is to minimize the variational bound on the negative log-likelihood, which can be simplified to a mean squared error loss between the predicted noise and the actual noise \cite{ho2020denoising}:
\begin{equation}
\mathcal{L}(\theta) = \mathbb{E}_{t, {x}_0, \epsilon} \left[ \|\epsilon - \epsilon_\theta({x}_t, t)\|^2 \right]
\end{equation}
where $\epsilon \sim \mathcal{N}(0, \mathbf{I})$ is the Gaussian noise, and ${x}_t$ is the noisy data generated during the forward process.

In the context of ITS, DDPM-based models have been employed for multiple traffic tasks such as trajectory prediction, traffic scenario generation, and spatio-temporal data completion. The simplicity of their training objective and compatibility with structured conditioning makes them a strong foundation for generative modeling in transportation domains.
\vspace{5px}
\subsubsection{\textbf{Noise Conditioned Score Networks (NCSNs)}} \

NCSNs \cite{song2019generative} are a class of score-based generative models that estimate the data distribution's score function. Instead of explicitly modeling the reverse diffusion process, NCSNs learn the gradient of the log-density of the data distribution at various noise levels via score matching \cite{hyvarinen2005estimation}, and subsequently generate samples via Langevin dynamics \cite{neal2012mcmc}.

\textbf{Score Matching.} Given an unknown data distribution $p_{data}(x)$, the score function of the data density $p(x)$ is defined as $\nabla_{x} \log p(x)$. The score network $\mathbf{s}_\theta$, a neural network parameterized by $\theta$, is trained to estimate the score function $\nabla_{x} \log p(x)$. When the data distribution is unknown, score estimation can be performed using sliced score matching \cite{song2020sliced} or denoising score matching \cite{vincent2011connection}. In NCSNs \cite{song2019generative}, denoising score matching is adopted, wherein data are perturbed with multiple levels of Gaussian noise. Specifically, the noise distribution is pre-specified as $q_{\sigma}(\Tilde{x} | x) = \mathcal{N}(\Tilde{x} | x, \sigma^2\mathbf{I})$, and the gradient of the log-likelihood with respect to the noisy data is given by $\nabla_{\Tilde{x}} \log q_\sigma(\Tilde{x} | x) = -(\Tilde{x} - x) / \sigma^2$. Given a sequence noise scales $\sigma_1 < \sigma_2 < ... < \sigma_L$, the denoising score matching objective for all $\sigma \in \{\sigma_i\}^L_{i=1}$ is defined as:
\begin{equation}
\mathcal{L} = \frac{1}{L}\sum_{i=1}^{L}\lambda(\sigma_i) \mathbb{E}_{p(x)} \mathbb{E}_{\Tilde{x} \sim q_{\sigma_i}(\Tilde{x} | x) } \left[ \left\|\mathbf{s}_\theta(\Tilde{x}, \sigma_i) + \frac{\Tilde{x} - x}{\sigma_i^2} \right\|_2^2 \right]
\end{equation}
where $\Tilde{x}$ is a noised version of $x$, and $\lambda(\sigma_i)$ is a weighting function depending on $\sigma_i$.

\textbf{Langevin Dynamics.} To generate samples, NCSNs employ annealed Langevin dynamics, starting with large noise levels and gradually annealing down to lower noise levels.  At each noise level, Langevin dynamics is iteratively applied using the learned score function to progressively recover the original data distribution. The update rule for Langevin dynamics is given by:
\begin{equation}
\Tilde{x}_t = \Tilde{x}_{t-1} + \frac{\alpha_i}{2} \mathbf{s}_\theta(\Tilde{x}_{t-1}, \sigma_i) + \sqrt{\alpha_i}\mathcal{N}(0, \mathbf{I})
\end{equation}
where $\alpha_i = \epsilon \cdot \sigma_i^2 / \sigma_L^2$, and $ t \in [1, T]$. When $\alpha_i \rightarrow 0$ and $T \rightarrow \infty$,  the final generated sample converges to the original data distribution $p_{data}(x)$.

While NCSNs are less commonly used in traffic applications compared to DDPMs, their formulation provides theoretical insights into score-based generation and has inspired hybrid models for spatial-temporal graph generation in traffic flow analysis \cite{wen2023diffstg}.

\begin{figure*}
\centering
\includegraphics[width=\linewidth, trim= 0 0 0 0, clip]{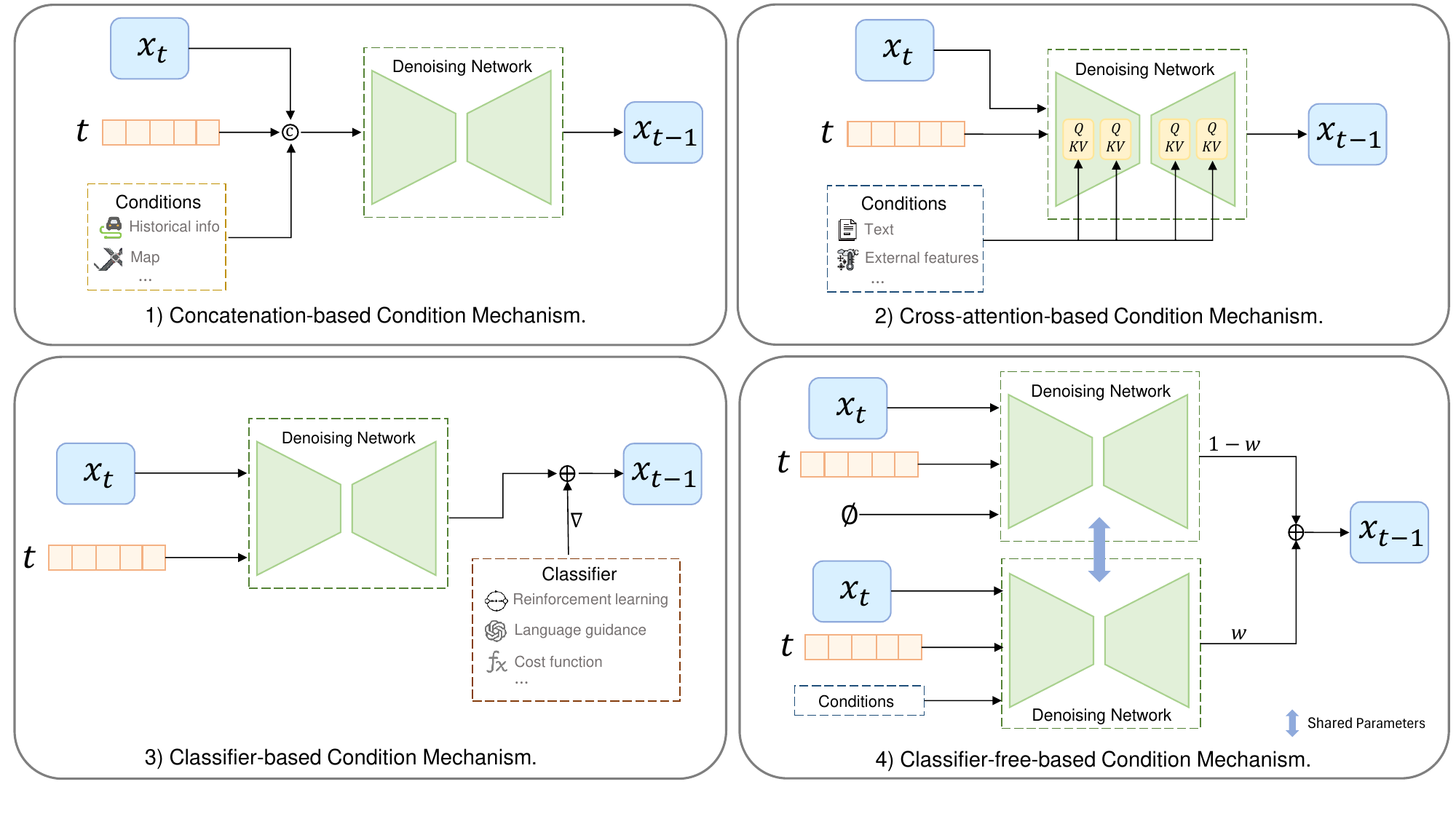}
\caption{Different condition mechanisms for diffusion models. (1) Concatenation-based mechanism directly incorporates conditions such as historical data and maps into the input. (2) Cross-attention-based mechanism integrates conditions like text and external features through cross-attention layers. (3) Classifier-based mechanism uses an external classifier to guide denoising based on conditions such as reinforcement learning or cost functions. (4) Classifier-free mechanism combines conditional and unconditional denoising models, balancing both with a weight parameter.}
\label{fig:condition}
\end{figure*}

\vspace{5px}
\subsubsection{\textbf{Stochastic Differential Equations (SDEs)}} \

SDEs \cite{song2020score} provide a continuous-time framework that unifies the concepts of DDPMs and NCSNs. Specifically, both the forward and reverse processes in these models are formulated as solutions to stochastic differential equations, with the reverse process requiring the estimation of score functions for noisy data distributions.

\textbf{Forward Process.} In the SDEs \cite{song2020score}, the forward process can be represented as the solution to an It\^{o} SDE \cite{ito1951stochastic}:
\begin{equation}
d{x} = \mathbf{f}({x}, t) dt + g(t) d\mathbf{w}
\end{equation}
where $\mathbf{f}(\cdot, t)$ denotes the drift coefficient of $x(t)$, $g(\cdot)$ represents the diffusion coefficient of $x(t)$, and $\mathbf{w}$ is a Brownian motion.

The forward processes in DDPMs and NCSNs can be regarded as discretizations of two different SDEs \cite{song2020score}. For DDPMs, the corresponding SDE is:
\begin{equation}
d{x} = -\frac{1}{2} \beta(t) x dt + \sqrt{\beta{t}} d\mathbf{w}
\end{equation}
whereas for NCSNs, the corresponding SDE is expressed as:
\begin{equation}
d{x} = \sqrt{\frac{d[\sigma^2(t)]}{dt}} d\mathbf{w}
\end{equation}

\textbf{Reverse Process.}
To generate samples, starting from samples of the standard Gaussian distribution $x(T)$ and reversing the process, the reverse-time SDE is solved \cite{anderson1982reverse}:
\begin{equation}
d{x} = [\mathbf{f}({x}, t) - g(t)^2 \nabla_{{x}} \log p_t({x})] dt + g(t) d\mathbf{\bar{w}}
\end{equation}
where $\mathbf{\bar{w}}$ is a Brownian motion with time flows backwards from $T$ to 0, and $dt$ is an infinitesimal negative timestep.

Similar to NCSNs, to estimate the score function $ \nabla_{{x}} \log p_t({x})$, we train a time-dependent score model $\mathbf{s}_\theta(x_{t}, t)$ by generalizing the score matching objective to continuous time. The objective function is given by:
\begin{equation}
\footnotesize
\mathcal{L}\!=\!\mathbb{E}_t\!\left\{\lambda(t)\mathbb{E}_{x(0)}\mathbb{E}_{x(t)|x(0)}\!\left[\left\|\mathbf{s}_\theta(x(t),t)\!-\!\nabla_{x(t)}\!\log p(x(t)\!\mid\!x(0)) \right\|_2^2 \right] \right\}
\end{equation}
where $t$ is uniformly sampled over the interval $[0, T]$, and $\lambda(t)$ is a positive weighting function.

The SDE-based formulation also provides a flexible framework for modeling traffic-related processes in continuous time, which is particularly advantageous in tasks involving irregularly sampled time series. For example, the CSDI model \cite{tashiro2021csdi} applies conditional score-based diffusion to time series imputation, interpolation, and forecasting, and achieves state-of-the-art performance on datasets with high missingness and non-uniform sampling intervals.

\subsection{Variants of Diffusion Models}
In this section, we introduce key variants of diffusion models, including conditional diffusion models and latent diffusion models (LDMs), which have significantly advanced the field of intelligent transportation systems. These models enhance the ability to generate realistic traffic data and offer flexibility and controllability in modeling complex traffic environments. By incorporating domain-specific information, such as historical data, traffic layouts, or external semantic features, conditional diffusion models enable the generation of more accurate and diverse traffic scenarios that reflect real-world conditions. Meanwhile, LDMs operate in a lower-dimensional latent space, facilitating faster training and inference times while maintaining the fidelity of generated outputs. Additionally, LDMs allow multi-modal conditions within the latent space. These capabilities make LDMs particularly useful for image-based, video-based, or text-involved traffic tasks. These advanced models demonstrate the potential of diffusion models to revolutionize intelligent transportation systems, providing powerful tools for traffic simulating, forecasting, and optimization in increasingly dynamic urban environments.

\subsubsection{Conditional Diffusion Models} \

The three types of standard diffusion models introduced above are unconditional, where the inputs are limited to the perturbed data $x_t$ and the diffusion step $t$. Conditional diffusion models, on the other hand, incorporate conditional information as an extra input, allowing for control over the generation process according to specific requirements. This capability makes them highly adaptable for various applications in intelligent transportation systems. Below, we focus on four primary conditioning mechanisms: concatenation-based, cross-attention-based, classifier-based, and classifier-free-based approaches. Concatenation-based methods are simple to implement but may struggle to capture complex relationships between the data and conditions. Cross-attention-based methods excel at modeling long-range dependencies and complex interactions with multi-modal conditioning, but they do not offer control over the strength of the conditions. Classifier-based approaches provide adjustable guidance through external classifiers but can be limited by the accuracy and generalization capability of the classifier. Classifier-free-based methods are flexible and do not require additional classifiers, but they often come with increased training costs. The visualization of these four conditioning mechanisms is shown in Fig.~\ref{fig:condition}.

\textbf{Concatenation-based.}
In concatenation-based mechanisms, the conditioning information is directly concatenated with the perturbed data $x_t$ or the diffusion step $t$, and then fed into the model for sample generation. This simple and effective method allows the model to leverage the conditioning information throughout the denoising process.
This approach is effective for ITS tasks like trajectory and traffic flow prediction, where historical data are concatenated with input for more accurate modeling \cite{gu2022stochastic, jiang2023motiondiffuser, chen2023equidiff}.

\textbf{Cross-attention-based.} 
Cross-attention-based conditional diffusion models integrate the cross-attention layers \cite{vaswani2017attention} into the denoising networks, enabling effective fusion of conditioning information during the denoising process and guiding the network to generate outputs aligned with the conditions. The cross-attention mechanism plays an important role in facilitating the interaction between the conditioning information and the noisy data, especially in scenarios where their relationship is complex or involves different modalities, such as text and images. Stable Diffusion \cite{rombach2022high} introduced a general-purpose conditioning mechanism based on cross-attention, enabling multi-modal conditional inputs, making diffusion models into powerful and flexible generators.
In ITS scenarios, cross-attention is commonly used when the condition (e.g., textual instruction) differs in modality from the input. For instance, Panacea \cite{wen2024panacea} applies cross-attention to align BEV features with textual prompts for scene generation.

\textbf{Classifier-based.}
The classifier-based mechanism incorporates conditions by using a task-related classifier to guide the diffusion sampling process, enabling controllable generation. Dhariwal and Nicho \cite{dhariwal2021diffusion} proposed a classifier-guidance approach, where an additional classifier $p_{\phi}(y|x_t, t)$ is trained on noisy data $x_t$ and the diffusion step $t$. The gradients of the guidance $\nabla_{x_t} \log p_{\phi}(y|x_t, t)$ are then used to guide the diffusion sampling process towards a specified class label $y$. Given a pre-trained diffusion model $p_\theta(x_t, t)$ and a pre-trained classifier $p_{\phi}(y|x_t, t)$, the diffusion sampling process is as follows:
\begin{equation}
x_{t-1} = \mathcal{N}(\mathbf{\mu}_\theta(x_t, t) + w \nabla_{x_t} \log p_{\phi}(y|x_t, t), \sigma^2 \mathbf{I})
\end{equation}
where $w$ is a hyperparameter controlling the strength of the guidance; as $w$ increases, the generated samples more closely adhere to the specified conditions.

This approach is commonly used in ITS tasks requiring controllable outputs, such as trajectory generation and traffic scenario simulation tasks, ensuring conformity to traffic rules and behavior constraints \cite{janner2022planning, zhong2023guided}.

\textbf{Classifier-free-based.}
The classifier-free mechanism combines unconditional and conditional diffusion models, achieving a balance between fidelity and diversity without the need to train a separate classifier. Additionally, it should be noted that the conditional diffusion model can employ either a concatenation mechanism or a cross-attention mechanism. In classifier-free diffusion guidance \cite{ho2022classifier}, the authors jointly train a conditional and an unconditional diffusion model, setting the condition $\textbf{c}$ to $\varnothing$ for the unconditional model. Then, a weighted average of the conditional and unconditional scores is used to estimate the score function:
\begin{equation}
\Tilde{\epsilon}_t = w\, \epsilon_\theta(x_t, t, \textbf{c}) + (1 - w)\, \epsilon_\theta(x_t, t, \varnothing)
\end{equation}
where $w$ is also a guidance scale.

This method is applied in ITS tasks like traffic scenario generation, where the goal is to maintain realistic outputs while allowing for diverse possibilities \cite{hu2023gaia}.

\textbf{Types of Conditions in ITS Applications.}
In intelligent transportation systems (ITS), diffusion models utilize a variety of conditions, which play critical roles in incorporating context, enforcing constraints, and improving overall model performance. We summarize the main types as follows:
\begin{itemize}
    \item \textbf{Sensor-based features:} Inputs from LiDAR or cameras \cite{ji2023ddp, wen2024panacea} provide rich environmental context, improving the accuracy of perception tasks such as segmentation and detection.
    \item \textbf{Spatial features:} Information such as road networks \cite{wei2024diff}, BEV features \cite{zou2024diffbev}, or traffic layouts \cite{li2023drivingdiffusion} encode spatial structure and constraints, enhancing realism and consistency in tasks like trajectory prediction and scenario simulation.
    \item \textbf{Temporal inputs:} Historical trajectories \cite{gu2022stochastic, jiang2023motiondiffuser}, departure times \cite{zhu2023difftraj}, and observation intervals provide motion context and support accurate future prediction and planning under uncertainty.
    \item \textbf{Semantic or symbolic conditions:} Text prompts \cite{wu2023diffumask, zhang2023chattraffc}, driving intents \cite{pronovost2023scenario}, traffic rules encoded via STL formulas \cite{zhong2023guided}, and behavior classes \cite{niedoba2024diffusion} enable high-level control, allowing the model to generate customized, diverse, or rule-compliant outputs.
\end{itemize}
These diverse conditions enhance the controllability, contextual awareness, and robustness of diffusion models across various ITS tasks, enhancing effectiveness across a wide range of ITS applications.

\begin{figure}[t]
    \centering
    \includegraphics [width=\linewidth, trim= 165 250 310 20, clip] {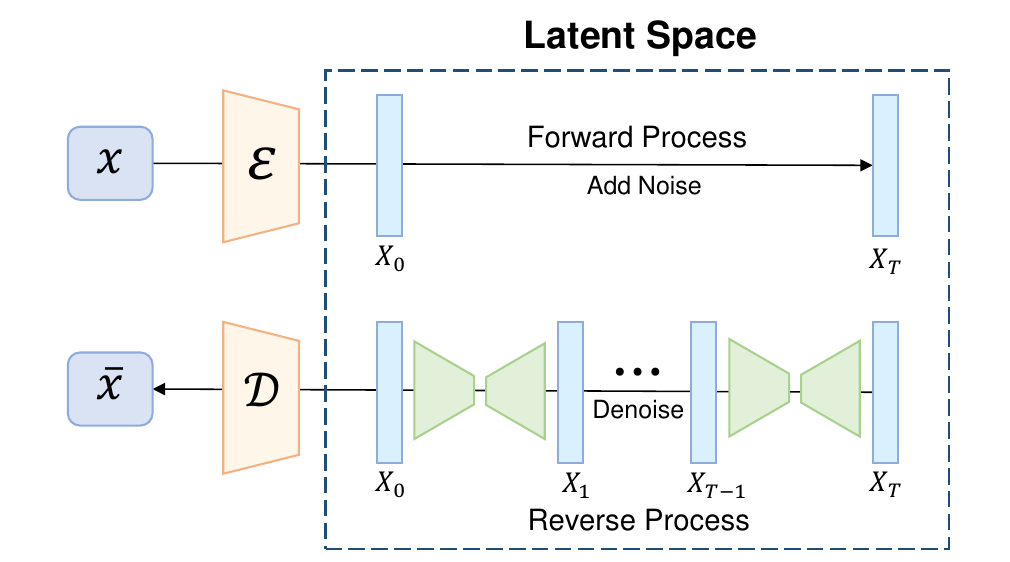}
    \caption{Illustration of latent diffusion models. Compared to standard diffusion models, they incorporate a pre-trained encoder $\mathcal{E}$ and decoder $\mathcal{D}$, with the diffusion and denoising processes operating in latent space rather than pixel or data space.}
    \label{fig:ldm}
\end{figure}

 \subsubsection{Latent Diffusion Models} \

The latent diffusion models (LDMs) \cite{rombach2022high} incorporate pre-trained perceptual compression models, VQGAN \cite{esser2021taming}, which consist of an encoder $\mathcal{E}$ and a decoder $\mathcal{D}$. As shown in Fig.~\ref{fig:ldm}, the input data (e.g., images or traffic states) are first mapped into a lower-dimensional latent representation $X = \mathcal{E}(x)$, where the diffusion process is applied. During training, Gaussian noise is gradually added to $X$ over several time steps, and a neural network $\epsilon_\theta$ is trained to predict and remove this noise. After the denoising process in latent space, the clean latent representation is decoded back to the data space via $\mathcal{D}(X)$. This latent-space approach significantly reduces computational overhead compared to pixel-space diffusion, allowing for more scalable and efficient training and inference while preserving generation fidelity. Building upon this architecture, Blattmann et al. \cite{blattmann2023align} extended LDMs to the temporal domain, proposing video latent diffusion models (VLDMs), which introduced temporal self-attention layers and adapted the autoencoder to handle video sequences.

LDMs have gained attention in intelligent transportation systems due to their ability to model complex traffic patterns and generate realistic traffic scenarios. This approach has proven particularly useful in simulating traffic flows \cite{zhang2023chattraffc}, predicting vehicle trajectories \cite{jiang2023motiondiffuser, balasubramanian2023scenediffusion, wang2024dragtraffic, pronovost2023scenario}, and enhancing autonomous driving systems through the generation of diverse and realistic traffic scenario data \cite{wang2024driving, yang2024generalized, wang2023drivedreamer, wen2024panacea,lu2023wovogen, ran2024towards}.

Moreover, LDMs are well-suited for handling multimodal traffic data and supporting controllable generation conditioned on various modalities. For example, by incorporating textual prompts, LDMs can generate traffic scenes that match specific descriptions \cite{wen2024panacea}, while conditioning on semantic road information allows for spatially consistent and context-aware traffic flow generation \cite{zheng2023diffuflow}. This controllability enables more flexibility in ITS applications, particularly in tasks that require alignment with high-level intent or environmental constraints.

\section{Challenges and Techniques}
\label{sec:challenge}
\begin{figure}[t]
    \centering
    \includegraphics [width=\linewidth, trim= 200 100 200 20, clip] {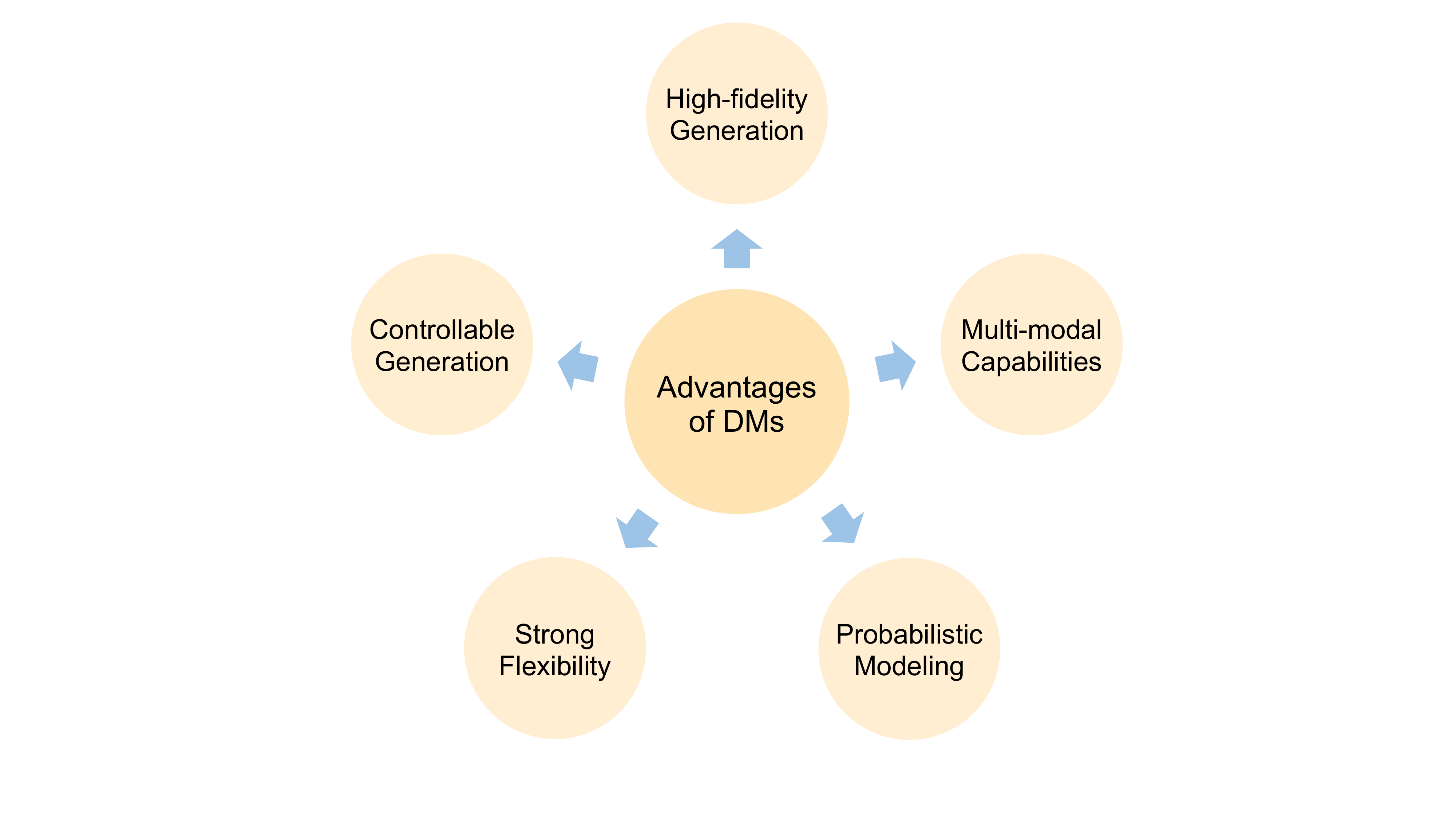}
    \caption{The advantages of diffusion models.}
    \label{fig:advantages}
\end{figure}

This section discusses key challenges in ITS and highlights why diffusion models, as a state-of-the-art generative approach, offer innovative solutions to these challenges. The complexity of traffic systems, combined with the inherent uncertainty and variability in traffic data, presents significant challenges for developing robust models. These challenges are further compounded by issues such as poor data quality, privacy concerns, and the need for scalable solutions that generalize effectively across different regions and traffic conditions. While various techniques have been developed to address these challenges, diffusion models have emerged as a promising approach due to their advantages: high-fidelity generation, controllable generation, strong flexibility, probabilistic modeling, and multi-modal capabilities. These strengths enhance the accuracy and robustness of ITS models, improving their applicability across diverse scenarios within the ITS field. As illustrated in Fig.~\ref{fig:challenge} and Fig.~\ref{fig:advantages}, the key challenges in ITS and the corresponding advantages of diffusion models are highlighted.

\subsection{Challenges in Intelligent Transportation Systems}
ITS is a sophisticated system that integrates advanced technologies and data analytics into transportation infrastructure and management to enhance the efficiency and safety of transportation networks \cite{veres2019deep, khalil2024advanced}. ITS encompasses a broad range of applications, including traffic prediction, autonomous driving, traffic simulation, and so on, all aimed at improving transportation services using large-scale traffic data and automated systems. However, several challenges affect the effectiveness and implementation of ITS:

  \begin{figure}[t]
    \centering
    \includegraphics [width=\linewidth, trim= 120 200 120 180, clip] {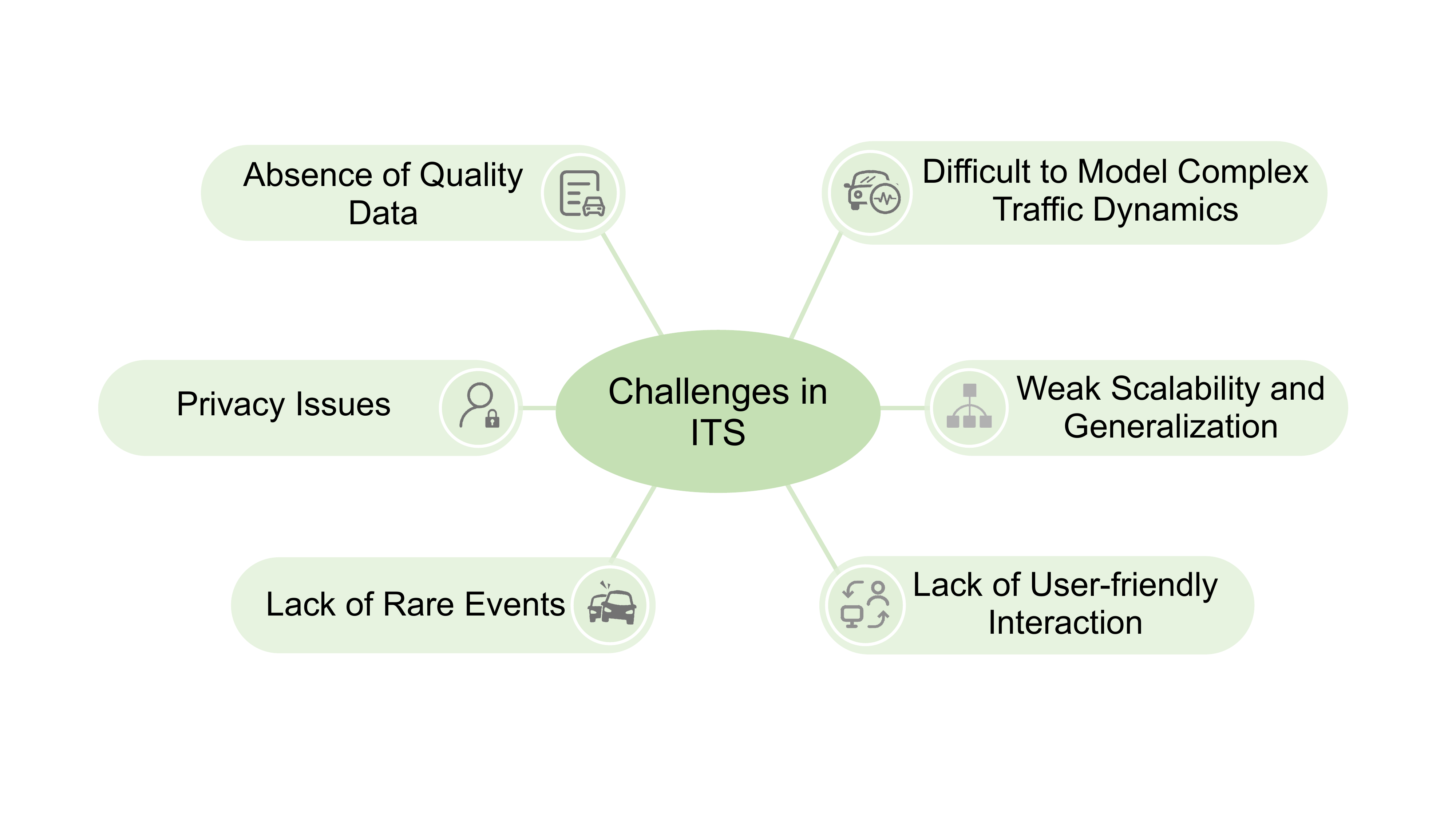}
    \caption{The challenges in intelligent transportation systems.}
    \label{fig:challenge}
\end{figure}

\begin{itemize}
    \item \textbf{Absence of Quality Data.} High-quality data are crucial for training reliable models, particularly in supervised learning approaches. However, real-world traffic data collected from traffic sensors, vehicle sensors, or GPS devices are often noisy, incomplete, or insufficient, limiting the ability to predict and simulate traffic conditions accurately.
    \item \textbf{Privacy Issues.} The collection of real-world traffic data from various sources, such as vehicle sensors, GPS devices, and surveillance cameras, raises significant privacy concerns. In particular, obtaining GPS data for traffic flow-related tasks is often challenging due to the need to protect personal and location information.
    \item \textbf{Lack of Rare Events.} Rare but critical events, such as accidents, sudden weather changes, or unexpected road blockages, are challenging to model due to their infrequency. This scarcity of data on such events makes it challenging to develop systems that can effectively handle and respond to these situations.
    \item \textbf{Difficult to Model Complex Traffic Dynamics.} Traffic systems are inherently complex, involving spatial and temporal dynamics at various scales and external factors such as holidays, weather conditions, and local events. Accurately modeling these dynamics and capturing the intricate relationships between different elements in the transportation network remains a challenge.
    \item \textbf{Weak Scalability and Generalization.} Many ITS solutions struggle to scale effectively or generalize across different regions and traffic conditions. Solutions that work well in one location may not perform as effectively in another due to variations in traffic patterns, and other local factors.
    \item \textbf{Lack of User-friendly Interaction.} Many current ITS interfaces and tools are difficult for users to navigate and use effectively. Improving user-friendly interaction is essential to ensure that users can easily understand and utilize the benefits of ITS technologies.
\end{itemize}

These challenges are common across multiple ITS tasks and highlight the need for more capable generative solutions. In the next section, we show how diffusion models are well-suited to address these issues.

\subsection{Advantages of Diffusion Models}
In ITS, various deep learning methods have been employed to address key challenges in various traffic tasks. For example, RNNs \cite{elman1990finding, hochreiter1997long} have proven effective in modeling temporal relationships in traffic data, and Transformers \cite{vaswani2017attention} are widely employed for multi-timestep traffic forecasting. Additionally, graph-based techniques such as Graph Neural Networks (GNNs) \cite{scarselli2008graph} and Graph Convolutional Networks (GCNs) \cite{kipf2016semi} have emerged as powerful tools for modeling traffic as graph structures, effectively capturing spatial interactions in transportation networks. However, these approaches often require large amounts of labeled data and tend to perform poorly with noise or incomplete data.

In contrast, generative models serve as flexible frameworks that can not only incorporate architectures such as CNNs, RNNs, and GNNs, enhancing their representational capacity, but are also effective at traffic data generation and imputation. 
However, generative models such as GANs \cite{goodfellow2014generative, creswell2018generative} often suffer from issues like mode collapse and unstable training. Mode collapse occurs when the model fails to capture the diversity of the data distribution, instead generating a narrow range of outputs, which limits the ability to simulate diverse traffic scenarios in ITS applications. Furthermore, GANs are notoriously difficult to train, requiring careful balancing between the generator and discriminator networks. This instability can be especially problematic in real-world traffic modeling tasks, where data is noisy, incomplete, or imbalanced. On the other hand, VAEs \cite{kingma2013auto, rezende2014stochastic} also exhibit limitations in their latent space expressiveness. While VAEs are more stable to train compared to GANs, they often produce lower-quality outputs due to their reliance on variational approximations. In ITS tasks, this can result in blurry or overly smoothed traffic scenarios, which do not accurately capture the intricate, high-dimensional characteristics of real-world traffic patterns. Moreover, VAEs often struggle to learn the complex dependencies in traffic data, such as spatial-temporal correlations and multi-agent interactions, which are crucial for generating realistic traffic scenarios. 

Recently, diffusion models have emerged as a promising class of generative models, offering several unique advantages that make them particularly well-suited for ITS applications:
\begin{itemize}
    \item \textbf{High-fidelity Generation.} Diffusion models have demonstrated the ability to generate high-quality and diverse outputs in traffic-related tasks. Compared to GANs and VAEs, diffusion models exhibit greater ease of training and superior generative capabilities \cite{dhariwal2021diffusion}. Moreover, the gradual denoising process inherent in diffusion models helps mitigate the challenges posed by noisy or incomplete real-world data, leading to more accurate traffic predictions and simulations. Additionally, by generating synthetic data, diffusion models can address traffic data privacy concerns, reducing the reliance on sensitive real-world data while maintaining the quality and reliability of generated scenarios.

    \item \textbf{Controllable Generation.} By incorporating task-related conditions—such as traffic layout, environmental factors, or textual instructions—conditional diffusion models can generate outputs aligned with specific goals or contexts. This controllability enables the simulation of customized traffic scenarios, including rare or unseen conditions, thereby helping address the lack of rare events.

    \item \textbf{Strong Flexibility.} Diffusion models can be flexibly integrated with other techniques, including GNNs, reinforcement learning, and even other generative models like GANs or VAEs. This enables them to capture intricate spatial-temporal dependencies and external influences, making them effective for modeling complex traffic dynamics. Their adaptability also improves transferability across regions, which contributes to better scalability and generalization.

    \item \textbf{Probabilistic Modeling.} The inherent probabilistic nature of diffusion models offers a robust framework for capturing uncertainty, variability, and stochasticity in traffic systems. This is particularly important for modeling complex traffic dynamics that involve multi-agent interactions, spatial-temporal dependencies, and external disturbances. By learning distributions rather than deterministic mappings, diffusion models provide a flexible way to represent the intricate behaviors observed in real-world traffic scenarios.

    \item \textbf{Multi-modal Capabilities.} Traffic data involves multiple modalities—such as trajectories, images, spatial-temporal graphs, BEV layouts, and textual descriptions. Diffusion models can effectively handle such data due to stepwise denoising mechanisms, flexible conditioning strategies, and latent space modeling. Latent diffusion models (LDMs) encode different modalities into compact latent spaces, allowing efficient and semantically aligned generation. For example, DrivingDiffusion \cite{li2023drivingdiffusion} fuses BEV maps and text prompts in latent space for scene generation, while DiffSTG \cite{wen2023diffstg} gradually denoising noisy spatio-temporal graphs to capture spatial and temporal dependencies through specialized architectures like UGnet. These capabilities help integrate heterogeneous data sources, and lay the foundation for more intuitive, user-friendly interaction through natural language, addressing the challenge of inaccessible ITS interfaces.
\end{itemize}

In the following sections, we will explore specific applications of diffusion models in the field of intelligent transportation systems, including autonomous driving, traffic simulation, traffic forecasting, and traffic safety. These applications will demonstrate how the advantages of diffusion models support their practical implementation in real-world traffic scenarios.

\section{Applications of Diffusion Models in ITS}

\subsection{Diffusion Models for Autonomous Driving}
\label{sec:autonomous}
Autonomous driving represents one of the most transformative aspects of ITS. The integration of autonomous vehicles (AVs) into ITS can drastically reduce traffic congestion, enhance safety, and improve the overall efficiency of transportation networks. However, achieving full autonomy in driving poses significant challenges due to the complex and dynamic nature of real-world driving environments, which are characterized by unpredictable events, diverse road conditions, and varying traffic behaviors \cite{zhao2023autonomous, chen2023end, teng2023motion}. Addressing these challenges requires advanced models capable of handling uncertainty, learning from vast amounts of data, and making real-time decisions in a safe and reliable manner. Diffusion models, with their ability to model complex distributions, refine data, and generate high-quality predictions, play a crucial role in advancing autonomous driving capabilities. 

\subsubsection{Perception} \

Perception in autonomous driving systems enables self-driving vehicles to sense and understand their environment \cite{grigorescu2020survey}. However, sensor data can be affected by weather, lighting, and other factors, which introduce noises and pose challenges for perception \cite{yurtsever2020survey}.  
With the rapid development of diffusion models in computer vision \cite{croitoru2023diffusion, rombach2022high, baranchuk2021label}, researchers are exploring their potential in enhancing autonomous driving perception. Diffusion models improve sensor data clarity and quality under various conditions \cite{zou2024diffbev, le2024diffuser, li2024light} and model uncertainty in perception \cite{nachkov2023diffusion, ji2023ddp}.
By leveraging these strengths, researchers aim to enhance perception tasks such as object detection, semantic segmentation, and object tracking, thereby contributing to safer and more reliable autonomous vehicles. The following section reviews recent progress in applying diffusion models to these tasks.

\textbf{Object Detection.} 
Object detection involves locating and sizing objects within an image by drawing bounding boxes around them \cite{yurtsever2020survey}. Recent advancements have introduced diffusion models to enhance detection accuracy. For example, Chen et al. \cite{chen2023diffusiondet} first redefined 2D object detection as a denoising diffusion process conditioned on the corresponding image, transforming noisy bounding boxes into precise object boxes. Additionally, Wang et al. \cite{wang2024detdiffusion} presented a pioneering framework that integrates diffusion models and perceptive models to enhance data generation quality and perception capabilities, using perception-aware attributes and loss to tailor images for specific perceptual criteria, boosting performance in object detection.

\textbf{Semantic Segmentation.} 
Semantic segmentation classifies each pixel in an image into predefined categories \cite{yurtsever2020survey}, and Bird's Eye View (BEV) perception is crucial for autonomous driving. Recent studies use diffusion models to improve BEV perception \cite{zou2024diffbev, nachkov2023diffusion, le2024diffuser}. Notably, Zhou et al. \cite{zou2024diffbev} first applied conditional diffusion models to denoise BEV features, enhancing segmentation and 3D object detection by refining object boundaries and shapes. Beyond BEV features, image features \cite{ji2023ddp} and text \cite{zhao2023unleashing} have been used as conditions. Ji et al. \cite{ji2023ddp} introduced DDP, a noise-to-map method guided by image features for visual perception, while Zhao et al. \cite{zhao2023unleashing} proposed VPD, leveraging pre-trained text-to-image diffusion models for semantic tasks by using textual inputs to guide image refinement. These approaches highlight the potential of diffusion models in enhancing semantic segmentation and visual perception.

\textbf{Object Tracking.} 
% \subsubsection{Object Tracking} \
Object tracking involves locating and maintaining the identity of objects in a video over time \cite{luo2021multiple}. Chen et al. \cite{chen2024delving} addressed trajectory length imbalance in multiple object tracking (MOT) by using Stationary and Dynamic Camera View Data Augmentation (SVA and DVA) and a Group Softmax module. The DVA, leveraging a conditional diffusion model, alters scene backgrounds to highlight pedestrian features, improving tracking performance. Additionally, Luo et al. \cite{luo2024diffusiontrack} proposed a noise-to-tracking framework that integrates detection and association through a denoising diffusion process, enabling consistency between detection and tracking. In contrast, Xie et al. \cite{xie2024diffusiontrack} introduced a point set-based denoising diffusion model for dynamic target localization, improving self-correction, simplifying post-processing, and enabling real-time tracking.

\textbf{Perception Data Generation.} 
% \subsubsection{Perception Data Generation} \
Recent advancements \cite{wu2023diffumask, wu2023datasetdm} have shown the effectiveness of diffusion models in synthesizing images and their corresponding annotations. Wu et al. \cite{wu2023diffumask} focused on semantic segmentation, using a text-guided pre-trained diffusion model to generate synthetic images with pixel-level semantic mask annotations. Building on this, Wu et al. \cite{wu2023datasetdm} introduced a dataset generation model that utilizes pre-trained diffusion models to produce diverse perception annotations, emphasizing a unified perception decoder that requires minimal human-labeled data to generate high-fidelity images paired with annotations such as depth, segmentation, and human pose estimation.

3D point cloud data has also seen significant progress with generative modeling. Luo et al. \cite{luo2021diffusion} proposed a generative model treating 3D point cloud generation as a reverse diffusion process, demonstrating flexibility and robustness in producing high-quality 3D point clouds. Following this work, Sun et al. \cite{sun2022pointdp} extended this work, addressing 3D point cloud recognition vulnerability to adversarial attacks by using the diffusion model from \cite{luo2021diffusion} as a base for adversarial point cloud purification.

\subsubsection{Trajectory Prediction} \

Trajectory prediction in autonomous driving systems involves using past states of traffic participants in a given scene to forecast their future states \cite{huang2022survey}. 
Key challenges include uncertainty, multi-modality in behavior, complex interactions, and environmental factors like road geometry \cite{huang2022survey, huang2023multimodal}. Recent advancements highlight diffusion models as an effective approach, as they capture the uncertainty and multi-modality of human and driving behavior, while also integrating map information, constraints, and other relevant factors.

\textbf{Human Trajectory Prediction.} 
To address unstable training and unnatural trajectories in human trajectory prediction, Gu et al. \cite{gu2022stochastic} introduced Motion Indeterminacy Diffusion (MID), which balances prediction diversity and determinacy by using a reverse diffusion process. However, its 17-second runtime for 100 diffusion steps is impractical for real-time use. To improve efficiency, Mao et al. \cite{mao2023leapfrog} developed a trainable leapfrog initializer that bypasses multiple denoising steps, enabling real-time prediction. Later, Bae et al. \cite{bae2024singulartrajectory} proposed a unified model, SingularTrajectory, introducing an adaptive anchor function as a good initializer similar to \cite{mao2023leapfrog} and leveraging a diffusion-based predictor to enhance prototype paths through a cascaded denoising process. Additionally, Liu et al. \cite{liu2024intention} introduced a PriorNet module for estimating prior noise distribution, reducing diffusion steps and consequently cutting inference time by two-thirds. Another study is LADM \cite{lv2024learning} combined VAEs with diffusion models to refine trajectories in a low-dimensional space for more accurate, real-time predictions.

Instantaneous trajectory prediction presents another challenge in human trajectory prediction due to the need for accurate predictions based on very limited observational data \cite{sun2022human}. Li et al. \cite{li2024bcdiff} addressed this challenge by utilizing bidirectional diffusion models to generate unobserved historical and future trajectories, incorporating a gate mechanism to balance the observed and predicted data.

\textbf{Vehicle Trajectory Prediction.} 
Vehicle trajectories are often governed by physical rules and constraints. 
Several studies have integrated these constraints and rules as classifiers \cite{jiang2023motiondiffuser} or conditions \cite{balasubramanian2023scenediffusion} into diffusion models for more realistic predictions.
Jiang et al. \cite{jiang2023motiondiffuser} utilized PCA-based latent diffusion models for multi-agent motion prediction and introduced constrained sampling based on differentiable cost functions. Similarly, Westny et al. \cite{westny2024diffusion} integrated differential motion constraints into the diffusion model output, generating realistic future trajectories. Another work by Balasubramanian et al. \cite{balasubramanian2023scenediffusion} employed conditional latent diffusion models with temporal constraints to predict the motion of vehicles in a traffic scenario, while also providing an unconditional mode as a scene initializer. 

Additionally, other works have combined diffusion models with other network architectures. For example, Chen et al. \cite{chen2023equidiff} combined diffusion models with an equivariant transformer to fully leverage the geometric properties of trajectories. Yao et al. \cite{yao2023graph} extended the MID model \cite{gu2022stochastic} by incorporating GNNs to model interactions between agents and road elements.

\subsubsection{Planning and Decision-making} \

In autonomous driving systems, planning and decision-making are crucial components. Planning entails generating a safe and comfortable trajectory based on the vehicle's current state, and environmental information \cite{teng2023motion}, while decision-making selects the optimal high-level action based on the goal, environment, traffic rules \cite{badue2021self}. 
Diffusion models have shown promise in improving generalization and flexibly integrating with other algorithms. They offer robust generalization to new environments with unseen obstacles \cite{carvalho2023motion, wang2024driving}, which is essential for dynamic environments.
Since the autonomous vehicle is a specialized form of robotics, we examine the topic within both the robotics and autonomous driving fields.

\textbf{Planning and Decision-making in Robotics.}
Diffusion models can flexibly combine with motion-planning approaches, such as reinforcement learning (RL) \cite{janner2022planning} or trajectory optimization algorithms \cite{carvalho2023motion}. Specifically, Janner et al. \cite{janner2022planning} proposed the Diffuser model, which combines RL with classifier-guided diffusion models \cite{dhariwal2021diffusion} to improve planning and decision-making processes. In contrast, Decision Diffuser \cite{ajay2022conditional}, employed classifier-free diffusion models \cite{ho2022classifier} to generate a sequence of future states, conditioning on rewards, various constraints, and behavior skills, eliminating the need for a separate classifier. A different approach was presented by Carvalho et al. \cite{carvalho2023motion}, proposed using learned diffusion priors to initialize an optimization-based motion planner, improving both the speed and diversity of trajectory planning.

\textbf{Planning and Decision-making in AVs.}  
Diffusion models have been employed to optimize the planning process in autonomous driving. Yang et al. \cite{yang2024diffusion} first combined gradient-free evolutionary search with diffusion models to enhance planning for autonomous driving. Unlike conventional methods that use naive Gaussian perturbations, this approach leverages a truncated diffusion-denoising process to mutate trajectories in the evolutionary search process, ensuring that the resulting mutations remain within the data manifold.  

Other studies have applied diffusion models to generate out-of-distribution driving scenarios, enhancing planning performance \cite{wang2024driving, chen2024human, chen2024dynamic}. For instance, Wang et al. \cite{wang2024driving} used diffusion models to generate multi-view future state videos, improving event prediction and risk assessment for safer planning. GenAD \cite{yang2024generalized} can generalize across diverse datasets and adapt to various tasks, including language-conditioned and action-conditioned prediction.

RL has seen widespread application in planning and decision-making for autonomous driving \cite{kiran2021deep, aradi2020survey}. Recent advancements have incorporated diffusion models to improve sampling efficiency \cite{zhu2023diffusion}. For example, Wang et al. \cite{wang2022diffusion} introduced Diffusion-QL, which integrates a conditional diffusion model as the policy and combines it with Q-learning, while Liu et al. \cite{liu2024ddm} employed conditional diffusion models as the actor in an Actor-Critic decision-making framework, facilitating policy exploration and learning.

\subsection{Diffusion Models for Traffic Simulation}
\label{sec:simulation}
Traffic simulation is essential for developing and testing intelligent transportation systems, enabling the modeling and analysis of traffic behavior and interactions \cite{nguyen2021overview, yan2023survey}. Traditional methods, such as rule-based or data-driven models, often fail to capture real-world traffic complexity and lack controllability to generate diverse, customizable scenarios needed for safety-critical testing \cite{chen2024data, ding2023survey}. Furthermore, traffic data is often unavailable or suffers from privacy concerns, posing additional challenges for data-driven traffic simulations.

Diffusion models, as generative models, offer a promising solution by learning traffic patterns and generating high-fidelity simulations that closely resemble real-world scenarios. They also provide greater controllability, allowing customization of traffic scenarios, driving trajectories, and traffic flows based on specific conditions or guidance.

\subsubsection{Traffic Trajectory Generation} \

Traffic trajectory generation is crucial for developing and testing intelligent transportation systems, creating realistic paths for vehicles and pedestrians. Traditional heuristic-based models \cite{lopez2018microscopic} follow specific traffic rules but fail to capture real-world complexity, while data-driven approaches \cite{chen2024data} generate more human-like behaviors but lack controllability. Diffusion models excel in modeling traffic patterns and offer enhanced controllability through guidance mechanisms, making them ideal for realistic and flexible trajectory generation.

Recent studies have enhanced controllability using classifiers, such as Signal Temporal Logic (STL) \cite{zhong2023guided}, RL reward-based classifiers \cite{xie2024advdiffuser}, and language-based classifiers \cite{zhong2023language}. Zhong et al. \cite{zhong2023guided} introduced STL-based guidance for generating physically feasible and rule-compliant trajectories, later incorporating language instructions \cite{zhong2023language} to improve user-friendliness. Additionally, Wang et al. \cite{wang2024dragtraffic} enhanced user-friendliness and controllability by introducing user-defined context through the cross-attention mechanism. 

Meanwhile, recent research has increasingly focused on multi-agent joint trajectory generation, aiming to generate more interactive trajectories \cite{zhong2023guided, wang2024dragtraffic, niedoba2024diffusion, pronovost2023scenario, huang2024versatile, yang2024wcdt}. Notably, Niedoba et al. \cite{niedoba2024diffusion} combine classifier and classifier-free guidance diffusion models to generate joint trajectories for all agents in a traffic scene. Additionally, Pronovost et al. \cite{pronovost2023scenario}  integrated latent diffusion with object detection and trajectory regression to generate joint agent poses and trajectories.

In human trajectory simulation, Rempe et al. \cite{rempe2023trace} developed TRACE, a controllable pedestrian simulation system, which uses a trajectory diffusion model and a physics-based humanoid controller to generate paths constrained by user-defined factors.

\subsubsection{Traffic Scenario Generation.} \ 

Traffic scenario generation involves creating a temporal sequence of traffic scene elements that simulate the actions, interactions, and events of the participating agents within a driving environment \cite{riedmaier2020survey, ding2023survey}. It enhances the efficiency and safety of intelligent transportation systems by generating diverse, safety-critical scenarios. 
However, it faces two key challenges: Consistency, ensuring scenarios are temporally and multi-view coherent, and Controllability, allowing scenarios to align with specific conditions or objectives \cite{zhu2024sora, wen2024panacea}. Diffusion models address these challenges by modeling complex data distributions with high realism. They can be combined with techniques like cross-view attention, multi-stage generation, and post-processing to ensure consistency. Controllable diffusion models, such as ControlNet \cite{zhang2023adding}, can integrate multimodal inputs like layout, text, and segmentation to fine-tune models like Stable Diffusion \cite{rombach2022high}, enhancing controllability in driving scenario generation.

With advances in image \cite{rombach2022high}, video generation \cite{blattmann2023align, ho2022video, ho2022imagen}, and world models \cite{ha2018recurrent, zhu2024sora}, diffusion models offer a robust framework for generating high-quality, consistent, and controllable traffic scenarios. This section will review advancements in traffic scenario generation from image-based and point cloud-based approaches.

\textbf{Image-based Driving Scenario Generation.}  
Recent advancements in diffusion models have made significant progress in generating realistic and controllable image-based driving scenarios. For example, Harvey et al. \cite{harvey2022flexible} introduced the Flexible Diffusion Model (FDM), which optimizes frame sampling schedules and handles long-range temporal dependencies. Hu et al. \cite{hu2023gaia} combined a video diffusion decoder with a world model to create high-fidelity driving scenarios, offering fine-grained control through action and language conditioning. Similarly, DriveDreamer \cite{wang2023drivedreamer} focused on generating high-quality, controllable driving videos and policies that align with real-world traffic structures, while Zhao et al. \cite{zhao2024drivedreamer} proposed the DriveDreamer-2 framework, which leverages the power of finetuned-LLMs \cite{mao2023gpt, peng2024lc} to to translate user descriptions into agent trajectories and HD maps.

For multi-view driving video generation, Wen et al. \cite{wen2024panacea} used a pre-trained diffusion model and 4D attention mechanism to generate temporally consistent multi-view videos. Li et al. \cite{li2023drivingdiffusion} developed DrivingDiffusion to create spatially and temporally consistent urban driving videos. Another important work is WoVoGen \cite{lu2023wovogen}, addressed intra-world consistency in multi-camera street-view generation using 4D world volumes. Furthermore, these approaches \cite{wen2024panacea, li2023drivingdiffusion, lu2023wovogen} employed the ControlNet \cite{zhang2023adding} for fine-grained control, conditioned on BEV sequences, 3D layouts, or world volume-aware image features.

\textbf{Point Cloud-based Driving Scenario Generation.}
The generation of realistic driving scenarios from point cloud data has gained attention for traffic simulation \cite{ran2024towards, zhang2023learning, zyrianov2024lidardm}. Notably, Ran et al. \cite{ran2024towards} concentrated focused on generating realistic LiDAR scenes by incorporating geometric priors and using the pre-trained CLIP model \cite{radford2021learning} for controllability with text conditions. Zhang et al. \cite{zhang2023learning} proposed Copilot4D, which uses VQVAE \cite{van2017neural} to tokenize point clouds and combines MaskGIT \cite{chang2022maskgit} with discrete diffusion models \cite{austin2021structured} for efficient decoding and denoising, improving point cloud-based scene forecasting.

\subsubsection{Traffic Flow Generation} \

Traffic flow generation creates synthetic data to model vehicle or pedestrian movement across specific regions within a transportation network \cite{yan2023survey}. 
These synthetic data are crucial for macroscopic simulations \cite{nguyen2021overview}, as modeling real-world human mobility trajectories often suffers from privacy concerns. However, it presents challenges such as the non-independent nature of trajectories and external factors like traffic conditions and events. Diffusion models are well-suited for handling stochasticity and uncertainty in traffic flow generation, and can be combined with GCNs, RNNs, and attention mechanisms to model spatiotemporal dependencies. These models can also generate traffic patterns based on inputs like text, road networks, and external factors.

Recent advancements include DiffSTG \cite{wen2023diffstg} and ChatTraffic \cite{zhang2023chattraffc}, which use GCNs to model spatiotemporal dependencies, while TimeGrad \cite{rasul2021autoregressive} uses RNNs, and CSDI \cite{tashiro2021csdi} and STPP \cite{yuan2023spatio} use attention mechanisms. In contrast, Zhou et al. \cite{zhou2023towards} proposed the KSTDiff, which leverages an urban knowledge graph (UKG) to capture urban flow and a volume estimator that integrates region-specific features to guide the diffusion model's sampling across different regions. Notably, ChatTraffic \cite{zhang2023chattraffc} also presented the first text-to-traffic generation framework, using BERT \cite{devlin2018bert} for text embedding to guide traffic flow generation.

GPS trajectory generation \cite{zhu2023difftraj, wei2024diff} has also gained attention. Zhu et al. \cite{zhu2023difftraj} proposed Traj-UNet for controlled GPS trajectory generation based on trip region and departure time. Diff-RNTraj \cite{wei2024diff} generates trajectories conditioned on the road network. 

Additionally, Rong et al. \cite{rong2023complexity} developed a cascaded graph denoising diffusion method for generating region-level origin-destination flows by modeling the topology structure and mobility flows in new cities.

\subsection{Diffusion Models for Traffic Forecasting}
\label{sec:traffic_forecasting}
Traffic forecasting is crucial for optimizing flow, reducing congestion, and improving transportation efficiency. It involves predicting future conditions like flow rates and travel times from historical data. However, challenges arise due to the complexity of transportation networks and data quality issues \cite{jiang2022graph}.

Recent advancements in traffic forecasting increasingly leverage diffusion models to tackle these challenges. These models effectively capture the dynamic nature of traffic systems and account for uncertainties and noise in traffic data, making them well-suited for handling incomplete or imperfect datasets. As a result, diffusion models are becoming more widely applied in tasks like traffic flow prediction and travel time estimation.

\subsubsection{Traffic Flow Forecasting} \

Traffic flow forecasting predicts future traffic states, such as vehicle speeds, density, and flow rates, based on historical data \cite{jiang2022graph}. However, challenges arise from uncertainties in flow distributions, complex external factors, and noisy, incomplete data. Recent advancements have used diffusion models to address these issues by recovering traffic data \cite{zheng2024recovering}, capturing spatial-temporal dependencies, and handling uncertainties \cite{wen2023diffstg}.

Graph-based approaches effectively capture spatial correlations in traffic networks \cite{ye2020build, li2024survey}. Wen et al. \cite{wen2023diffstg} proposed UGnet, a GCN-based network that captures multi-scale temporal dependencies and spatial correlations. To reduce computational costs, Lin et al. \cite{lin2024specstg} introduced fast spectral graph convolution.

Diffusion models have also been used for fine-grained traffic flow inference from noisy and incomplete data. Zheng et al. \cite{zheng2023diffuflow} developed a transformer-based network to capture spatial-temporal dependencies and external factors, while Xu et al. \cite{xu2023diffusion} used relaxed structural constraints for flow map and external factor learning. Additionally, Lablack et al. \cite{lablack2023long} proposed a vectorized state space module to decompose the historical signal of an ego-graph into the frequency domain, thereby reducing the impact of noise and data imperfections present in real-world traffic data.

Lastly, Chi et al. \cite{chi2023difforecast} transformed traffic flow forecasting into a conditional image generation problem by introducing a space-time image that incorporates physical traffic variables.
 
\subsubsection{Travel Time Estimation} \

Origin-Destination (OD) travel time estimation predicts the time required to travel between a starting point and a destination, but variability in travel times due to factors like traffic and route choices makes accurate prediction challenging \cite{mori2015review}. Multiple historical trajectories with different travel times can complicate predictions, especially when outlier trajectories are present. To address this, Lin et al. \cite{lin2023origin} proposed a conditional diffusion model for OD travel time estimation. This model uses a pixelated trajectory representation conditioned on origin, destination, and departure time (ODT) to capture correlations between OD pairs and historical patterns, helping filter out outlier trajectories.

\subsection{Diffusion Models for Traffic Safety}
\label{sec:traffic_safety}
Traffic safety is crucial in intelligent transportation systems, aiming to minimizing the risks associated with vehicular travel and reducing the frequency and severity of traffic accidents \cite{goniewicz2016road}. Recent advancements in diffusion models have improved traffic safety by generating high-quality, customizable samples conditioned on text descriptions, addressing the challenge of limited traffic accident or anomaly data. These models have been effectively applied to traffic anomaly detection and accident prevention, enhancing the safety and efficiency of transportation systems.

\subsubsection{Traffic Anomaly Detection} \

Traffic anomaly detection identifies irregular patterns in traffic data, such as unusual vehicle activity or accidents, which is crucial for traffic management and safety.
However, challenges arise due to limited labeled anomaly data and difficulty in defining normal versus abnormal patterns \cite{santhosh2020anomaly, kong2024mobile}. Diffusion models offer a promising solution for traffic anomaly detection, as anomalous events often exhibit a level of randomness and uncertainty that are inherently similar to the diffusion process. By leveraging diffusion models to reconstruct normal traffic patterns from Gaussian noise, researchers can effectively identify samples that deviate from these normal patterns.

Li et al. \cite{li2024difftad} formalized vehicle trajectory anomaly detection as a noisy-to-normal process, using diffusion models to reconstruct near-normal trajectories and identify anomalies. Similarly, Yan et al. \cite{yan2023feature} employed diffusion models to learn normal motion and appearance features for video anomaly detection using denoising modules.

\subsubsection{Traffic Accident Prevention} \

Traffic accident prevention requires a deep understanding of accident causality and then designing strategies to reduce their likelihood. A significant challenge in this field is the lack of a large-scale and long-tailed accident dataset \cite{fang2023vision}, which limits the ability to develop comprehensive and effective accident prevention. With their powerful, controllable generation capabilities, diffusion models have emerged as a promising tool to overcome these challenges. 

Recent advancements in diffusion models have enabled innovative applications in accident analysis and prevention. For example, Fang et al. \cite{fang2024abductive} used an abductive CLIP model within Object-Centric Video Diffusion (OAVD) to identify cause-effect chains of accidents. By generating video frames conditioned on text descriptions like accident causes and prevention strategies, this approach helps visualize potential accident outcomes and improve prevention efforts.

\begin{figure}[ht]
    \centering
    \includegraphics [width=\linewidth, trim= 40 200 50 100, clip] {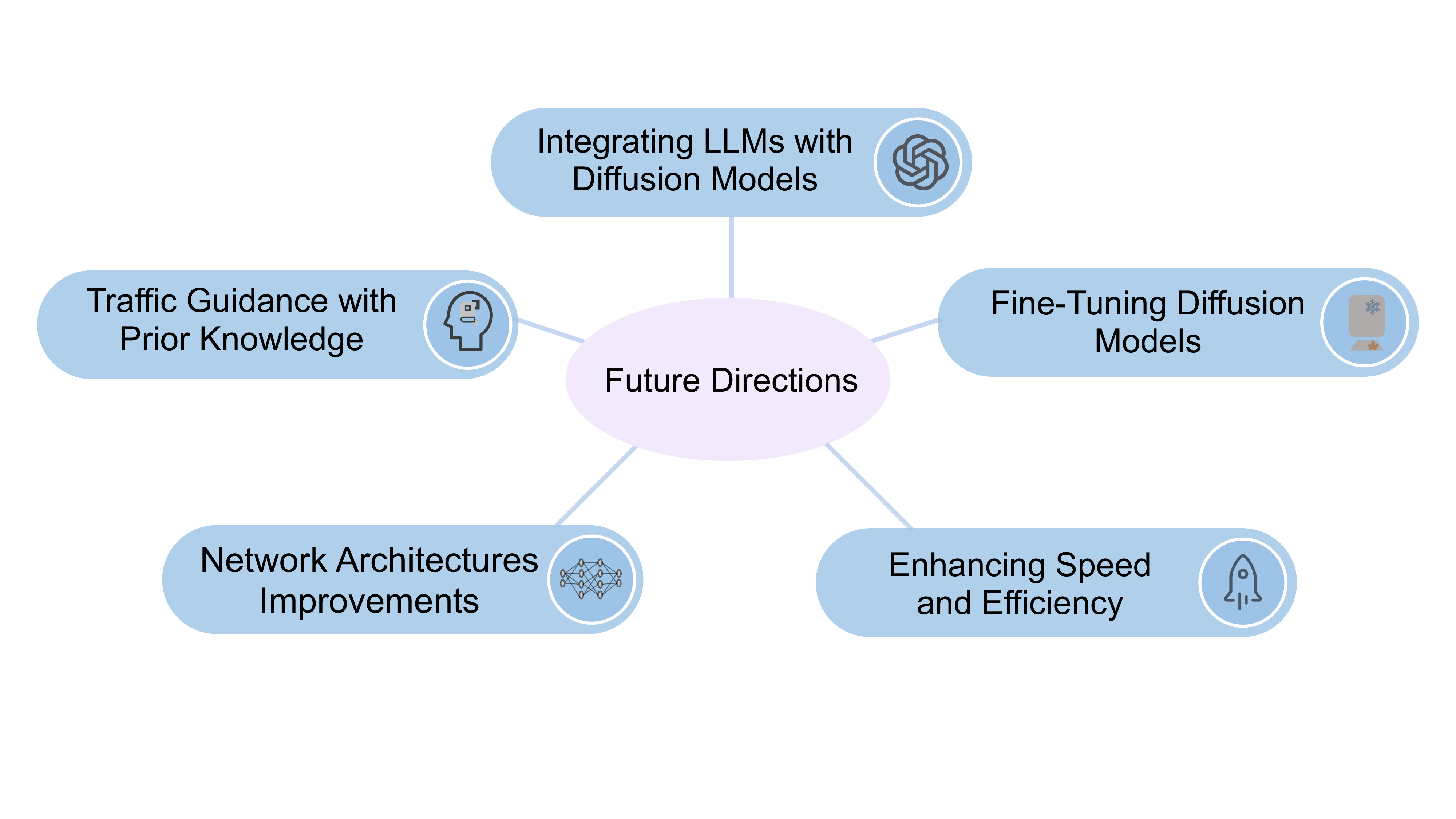}
    \caption{Future research directions for diffusion models in intelligent transportation systems.}
    \label{fig:future}
\end{figure}

\section{Future Directions}
\label{sec:future}
As diffusion models continue to evolve, their potential to address complex challenges in ITS becomes increasingly evident. However, several critical areas require further investigation and innovation to fully realize their capabilities. This section outlines key research directions for diffusion models in ITS that are worthy of further exploration, as shown in Fig.~\ref{fig:future}.

\subsection{Integrating LLMs with Diffusion Models}
The integration of LLMs and diffusion models represents a promising new direction in ITS. Previous works, such as \cite{wen2024panacea, li2023drivingdiffusion, ran2024towards}, have primarily relied on pre-trained CLIP \cite{radford2021learning} to encode textual information and generate outputs conditioned on these text feature representations. However, CLIP exhibits inherent limitations in processing long and complex sentences, which can negatively impact the quality of generated outputs. LLMs, with their strong capabilities in language understanding and knowledge-based reasoning, combined with the generative power of diffusion models, offer a compelling opportunity for enhanced performance. Recent studies, including MiniGPT-5 \cite{zheng2023minigpt}, which utilizes ``generative vokens" to bridge LLMs and diffusion models, and EasyGen \cite{zhao2023making}, which integrates these models via a projection layer, have demonstrated the potential for producing more realistic and reasonable outputs. 
Building on these advancements, the integration of large-scale foundational models, such as LLMs, with diffusion models for various ITS tasks presents a highly promising and emerging direction. In particular, in the field of traffic simulation, LLMs can facilitate semantic comprehension, reasoning, and automated decision-making, thereby enabling the generation of more realistic, contextually accurate driving scenarios \cite{zhang2024chatscene, zhong2023language}. Additionally, combining LLMs with diffusion models offers significant benefits as a user interface for ITS systems. The natural language processing capabilities of LLMs provide a more intuitive and accessible means for users to interact with these systems, allowing them to describe complex scenarios and receive tailored outputs without needing in-depth technical expertise. Moreover, this integration can enhance decision-making processes in smart cities, improving traffic flow analysis and resource allocation, due to the LLMs' capacity to holistically understand complex systems.

\subsection{Traffic Guidance with Prior Knowledge}
Traffic-related tasks often require reasoning that integrates both scenario-specific features and domain-specific knowledge. Rather than relying on a large, and computationally expensive diffusion model, the development of more efficient traffic guidance that incorporate prior knowledge about traffic systems can significantly enhance the generative process. Existing research has primarily focused on designing guidance to guide sampling in autonomous driving contexts, particularly in planning and decision-making. These guidance are often based on reinforcement learning techniques or cost functions grounded in traffic rules \cite{janner2022planning, carvalho2023motion}. Beyond autonomous driving, other domains within ITS, such as traffic flow prediction and traffic safety analysis, also rely heavily on domain knowledge. For instance, factors like the relationship between traffic flow, urban population density, public holidays, weather conditions, and landmark locations are critical for accurate traffic forecasting. By leveraging this extensive domain knowledge, task-specific guidance can be developed to improve the prediction of traffic patterns and congestion levels. Future research could focus on creating guidance that more effectively mine and utilize relevant prior knowledge for specific traffic-related tasks, thereby advancing the performance of diffusion models in these domains.

\subsection{Network Architectures Improvements}
The architectures of diffusion models present substantial opportunities for improvement. U-Net \cite{ronneberger2015u}, while demonstrating remarkable performance as a denoising network backbone across various traffic-related tasks and being combinable with methods such as GCNs to model spatial-temporal dependencies \cite{wen2023diffstg}, still has considerable potential for further optimization. Recent advancements in transformer-based denoising networks, such as DiT \cite{peebles2023scalable}, U-ViT \cite{bao2023all}, and their applications in diffusion models like Sora \cite{videoworldsimulators2024} and Stable Diffusion 3 \cite{esser2024scaling}, have gained significant attention. Transformer-based architectures excel in capturing long-range spatial-temporal relationships and offer greater scalability. Therefore, leveraging or refining transformer-based denoising networks holds significant potential for enhancing spatial-temporal-related traffic applications, such as traffic flow forecasting and traffic trajectory prediction. Furthermore, designing novel network architectures specifically tailored to particular tasks within intelligent transportation systems, as backbones for diffusion models, presents a promising direction for future research.

\subsection{Fine-Tuning Diffusion Models}
Large diffusion models, such as Stable Diffusion \cite{rombach2022high}, pre-trained on extensive image datasets, have demonstrated considerable promise across various domains. Fine-tuning these models on traffic-specific data or for traffic-related condition control can further enhance their applicability within ITS. Recent research has explored methods to fine-tune large pre-trained diffusion models for more fine-grained control. For example, ControlNet \cite{zhang2023adding} adds spatial conditioning controls to large and pre-trained diffusion models through efficient fine-tuning techniques. Similarly, T2I-Adapter \cite{mou2024t2i} learns simple and lightweight adapters to align internal knowledge in large diffusion models with external control signals. Building on these advancements, developing effective fine-tuning methods tailored to traffic data or traffic scenes holds the potential to significantly enhance the flexibility and control of these models in generating traffic-related outputs. These approaches promise to improve the models' utility in various ITS applications, particularly in traffic simulation and incident detection.

\subsection{Enhancing Speed and Efficiency}
Although diffusion models have demonstrated significant potential in generating high-quality results, their computational cost and slow inference speeds remain major bottlenecks. These limitations are particularly critical for real-time ITS applications and deployment on resource-constrained edge devices, such as those used in autonomous driving or real-time traffic monitoring systems. Recent advancements have focused on improving efficiency through various techniques \cite{song2020denoising, salimans2022progressive, song2023consistency, mao2023leapfrog, rombach2022high, xue2024accelerating}. For example, Denoising Diffusion Implicit Models (DDIM) \cite{song2020denoising} have introduced a non-Markovian process that reduces the number of sampling steps required. Progressive Distillation \cite{salimans2022progressive} improves speed by distilling slow models into faster ones that require fewer sampling steps. Consistency Models \cite{song2023consistency} allow for faster generation by mapping noise to data directly in a single step, bypassing the iterative denoising process. In addition, \cite{xue2024accelerating} designed an optimization problem to seek appropriate time steps to speed up the sampling process, while \cite{mao2023leapfrog} introduced a trainable leapfrog initializer to bypass multiple denoising steps. Nevertheless, further innovation is needed to meet the stringent requirements of latency and power consumption in real-world ITS scenarios. Future research should explore the development of more adaptive and lightweight network architectures that reduce model complexity while maintaining generation quality. Parallel or partially denoised sampling strategies could also significantly reduce inference time. Moreover, hybrid frameworks that combine diffusion models with faster, more deterministic generative methods (e.g., GANs or autoregressive models) may offer promising trade-offs between fidelity and speed, making them more suitable for deployment in edge environments and real-time decision-making tasks in ITS.

\section{Conclusion}
\label{sec:conclusion}
In this paper, we provide a comprehensive review of diffusion models in ITS. We outline the theoretical foundations of diffusion models, discuss their key variants, and demonstrate how they can effectively address the complex challenges of ITS. Our review also highlights the advantages of diffusion models, especially in handling multi-modal, noisy, and incomplete traffic data. By investigating their current applications in ITS domains, including autonomous driving, traffic simulation, traffic forecasting, and traffic safety, we highlight the versatility and potential of diffusion models in enhancing various aspects of ITS. Additionally, we summarize several key research directions that warrant further investigation, including the integration of other approaches and the development of more efficient and scalable diffusion models tailored to various traffic-related tasks. We hope this review encourages further interdisciplinary collaboration, paving the way for the continued evolution of diffusion models as a pivotal tool in future ITS.

\section*{Acknowledgments}
This study is supported by the National Natural Science Foundation of China under Grant 52302379, Guangdong Provincial Natural Science Foundation-General Project with Grant 2024A1515011790, Guangzhou Basic and Applied Basic Research Projects under Grants 2023A03J0106 and 2024A04J4290, Guangdong Province General Universities Youth Innovative Talents Project under Grant 2023KQNCX100, Guangzhou Municipal Science and Technology Project 2023A03J0011, Nansha District Key R\&D Project 2023ZD006.

\bibliographystyle{ieeetr}
\bibliography{reference}

\begin{thebibliography}{100}

\bibitem{wootton1995intelligent}
J.~Wootton, A.~Garcia-Ortiz, and S.~Amin, ``Intelligent transportation systems: a global perspective,'' {\em Mathematical and Computer Modelling}, vol.~22, no.~4-7, pp.~259--268, 1995.

\bibitem{veres2019deep}
M.~Veres and M.~Moussa, ``Deep learning for intelligent transportation systems: A survey of emerging trends,'' {\em IEEE Transactions on Intelligent Transportation Systems}, vol.~21, no.~8, pp.~3152--3168, 2019.

\bibitem{ye2020build}
J.~Ye, J.~Zhao, K.~Ye, and C.~Xu, ``How to build a graph-based deep learning architecture in traffic domain: A survey,'' {\em IEEE Transactions on Intelligent Transportation Systems}, vol.~23, no.~5, pp.~3904--3924, 2020.

\bibitem{li2024survey}
H.~Li, Y.~Zhao, Z.~Mao, Y.~Qin, Z.~Xiao, J.~Feng, Y.~Gu, W.~Ju, X.~Luo, and M.~Zhang, ``A survey on graph neural networks in intelligent transportation systems,'' {\em arXiv preprint arXiv:2401.00713}, 2024.

\bibitem{lin2023generative}
H.~Lin, Y.~Liu, S.~Li, and X.~Qu, ``How generative adversarial networks promote the development of intelligent transportation systems: A survey,'' {\em IEEE/CAA Journal of Automatica Sinica}, 2023.

\bibitem{boquet2020variational}
G.~Boquet, A.~Morell, J.~Serrano, and J.~L. Vicario, ``A variational autoencoder solution for road traffic forecasting systems: Missing data imputation, dimension reduction, model selection and anomaly detection,'' {\em Transportation Research Part C: Emerging Technologies}, vol.~115, p.~102622, 2020.

\bibitem{croitoru2023diffusion}
F.-A. Croitoru, V.~Hondru, R.~T. Ionescu, and M.~Shah, ``Diffusion models in vision: A survey,'' {\em IEEE Transactions on Pattern Analysis and Machine Intelligence}, 2023.

\bibitem{videoworldsimulators2024}
T.~Brooks, B.~Peebles, C.~Holmes, W.~DePue, Y.~Guo, L.~Jing, D.~Schnurr, J.~Taylor, T.~Luhman, E.~Luhman, C.~Ng, R.~Wang, and A.~Ramesh, ``Video generation models as world simulators,'' 2024.

\bibitem{khalil2024advanced}
R.~A. Khalil, Z.~Safelnasr, N.~Yemane, M.~Kedir, A.~Shafiqurrahman, and N.~Saeed, ``Advanced learning technologies for intelligent transportation systems: Prospects and challenges,'' {\em IEEE Open Journal of Vehicular Technology}, 2024.

\bibitem{yan2023survey}
H.~Yan and Y.~Li, ``A survey of generative ai for intelligent transportation systems,'' {\em arXiv preprint arXiv:2312.08248}, 2023.

\bibitem{yang2023diffusion}
L.~Yang, Z.~Zhang, Y.~Song, S.~Hong, R.~Xu, Y.~Zhao, W.~Zhang, B.~Cui, and M.-H. Yang, ``Diffusion models: A comprehensive survey of methods and applications,'' {\em ACM Computing Surveys}, vol.~56, no.~4, pp.~1--39, 2023.

\bibitem{jiang2024survey}
R.~Jiang, G.-C. Zheng, T.~Li, T.-R. Yang, J.-D. Wang, and X.~Li, ``A survey of multimodal controllable diffusion models,'' {\em Journal of Computer Science and Technology}, vol.~39, no.~3, pp.~509--541, 2024.

\bibitem{yang2024survey}
Y.~Yang, M.~Jin, H.~Wen, C.~Zhang, Y.~Liang, L.~Ma, Y.~Wang, C.~Liu, B.~Yang, Z.~Xu, {\em et~al.}, ``A survey on diffusion models for time series and spatio-temporal data,'' {\em arXiv preprint arXiv:2404.18886}, 2024.

\bibitem{kazerouni2023diffusion}
A.~Kazerouni, E.~K. Aghdam, M.~Heidari, R.~Azad, M.~Fayyaz, I.~Hacihaliloglu, and D.~Merhof, ``Diffusion models in medical imaging: A comprehensive survey,'' {\em Medical Image Analysis}, vol.~88, p.~102846, 2023.

\bibitem{sohl2015deep}
J.~Sohl-Dickstein, E.~Weiss, N.~Maheswaranathan, and S.~Ganguli, ``Deep unsupervised learning using nonequilibrium thermodynamics,'' in {\em International Conference on Machine Learning}, pp.~2256--2265, 2015.

\bibitem{ho2020denoising}
J.~Ho, A.~Jain, and P.~Abbeel, ``Denoising diffusion probabilistic models,'' {\em Advances in Neural Information Processing Systems}, vol.~33, pp.~6840--6851, 2020.

\bibitem{ronneberger2015u}
O.~Ronneberger, P.~Fischer, and T.~Brox, ``U-net: Convolutional networks for biomedical image segmentation,'' in {\em International Conference on Medical Image Computing and Computer Assisted Intervention}, pp.~234--241, 2015.

\bibitem{song2019generative}
Y.~Song and S.~Ermon, ``Generative modeling by estimating gradients of the data distribution,'' {\em Advances in Neural Information Processing Systems}, vol.~32, 2019.

\bibitem{hyvarinen2005estimation}
A.~Hyv{\"a}rinen and P.~Dayan, ``Estimation of non-normalized statistical models by score matching.,'' {\em Journal of Machine Learning Research}, vol.~6, no.~4, 2005.

\bibitem{neal2012mcmc}
R.~M. Neal, ``Mcmc using hamiltonian dynamics,'' {\em arXiv preprint arXiv:1206.1901}, 2012.

\bibitem{song2020sliced}
Y.~Song, S.~Garg, J.~Shi, and S.~Ermon, ``Sliced score matching: A scalable approach to density and score estimation,'' in {\em Uncertainty in Artificial Intelligence}, pp.~574--584, 2020.

\bibitem{vincent2011connection}
P.~Vincent, ``A connection between score matching and denoising autoencoders,'' {\em Neural Computation}, vol.~23, no.~7, pp.~1661--1674, 2011.

\bibitem{wen2023diffstg}
H.~Wen, Y.~Lin, Y.~Xia, H.~Wan, Q.~Wen, R.~Zimmermann, and Y.~Liang, ``Diffstg: Probabilistic spatio-temporal graph forecasting with denoising diffusion models,'' in {\em Proceedings of the 31st ACM International Conference on Advances in Geographic Information Systems}, pp.~1--12, 2023.

\bibitem{song2020score}
Y.~Song, J.~Sohl-Dickstein, D.~P. Kingma, A.~Kumar, S.~Ermon, and B.~Poole, ``Score-based generative modeling through stochastic differential equations,'' {\em arXiv preprint arXiv:2011.13456}, 2020.

\bibitem{ito1951stochastic}
K.~It{\^o}, {\em On stochastic differential equations}.
\newblock No.~4, 1951.

\bibitem{anderson1982reverse}
B.~D. Anderson, ``Reverse-time diffusion equation models,'' {\em Stochastic Processes and their Applications}, vol.~12, no.~3, pp.~313--326, 1982.

\bibitem{tashiro2021csdi}
Y.~Tashiro, J.~Song, Y.~Song, and S.~Ermon, ``Csdi: Conditional score-based diffusion models for probabilistic time series imputation,'' {\em Advances in Neural Information Processing Systems}, vol.~34, pp.~24804--24816, 2021.

\bibitem{gu2022stochastic}
T.~Gu, G.~Chen, J.~Li, C.~Lin, Y.~Rao, J.~Zhou, and J.~Lu, ``Stochastic trajectory prediction via motion indeterminacy diffusion,'' in {\em Proceedings of the IEEE Conference on Computer Vision and Pattern Recognition}, pp.~17113--17122, 2022.

\bibitem{jiang2023motiondiffuser}
C.~Jiang, A.~Cornman, C.~Park, B.~Sapp, Y.~Zhou, D.~Anguelov, {\em et~al.}, ``Motiondiffuser: Controllable multi-agent motion prediction using diffusion,'' in {\em Proceedings of the IEEE Conference on Computer Vision and Pattern Recognition}, pp.~9644--9653, 2023.

\bibitem{chen2023equidiff}
K.~Chen, X.~Chen, Z.~Yu, M.~Zhu, and H.~Yang, ``Equidiff: A conditional equivariant diffusion model for trajectory prediction,'' in {\em 2023 IEEE 26th International Conference on Intelligent Transportation Systems}, pp.~746--751, 2023.

\bibitem{vaswani2017attention}
A.~Vaswani, ``Attention is all you need,'' {\em Advances in Neural Information Processing Systems}, 2017.

\bibitem{rombach2022high}
R.~Rombach, A.~Blattmann, D.~Lorenz, P.~Esser, and B.~Ommer, ``High-resolution image synthesis with latent diffusion models,'' in {\em Proceedings of the IEEE conference on computer vision and pattern recognition}, pp.~10684--10695, 2022.

\bibitem{wen2024panacea}
Y.~Wen, Y.~Zhao, Y.~Liu, F.~Jia, Y.~Wang, C.~Luo, C.~Zhang, T.~Wang, X.~Sun, and X.~Zhang, ``Panacea: Panoramic and controllable video generation for autonomous driving,'' in {\em Proceedings of the IEEE Conference on Computer Vision and Pattern Recognition}, pp.~6902--6912, 2024.

\bibitem{dhariwal2021diffusion}
P.~Dhariwal and A.~Nichol, ``Diffusion models beat gans on image synthesis,'' {\em Advances in Neural Information Processing Systems}, vol.~34, pp.~8780--8794, 2021.

\bibitem{janner2022planning}
M.~Janner, Y.~Du, J.~B. Tenenbaum, and S.~Levine, ``Planning with diffusion for flexible behavior synthesis,'' {\em arXiv preprint arXiv:2205.09991}, 2022.

\bibitem{zhong2023guided}
Z.~Zhong, D.~Rempe, D.~Xu, Y.~Chen, S.~Veer, T.~Che, B.~Ray, and M.~Pavone, ``Guided conditional diffusion for controllable traffic simulation,'' in {\em 2023 IEEE International Conference on Robotics and Automation}, pp.~3560--3566, 2023.

\bibitem{ho2022classifier}
J.~Ho and T.~Salimans, ``Classifier-free diffusion guidance,'' {\em arXiv preprint arXiv:2207.12598}, 2022.

\bibitem{hu2023gaia}
A.~Hu, L.~Russell, H.~Yeo, Z.~Murez, G.~Fedoseev, A.~Kendall, J.~Shotton, and G.~Corrado, ``Gaia-1: A generative world model for autonomous driving,'' {\em arXiv preprint arXiv:2309.17080}, 2023.

\bibitem{ji2023ddp}
Y.~Ji, Z.~Chen, E.~Xie, L.~Hong, X.~Liu, Z.~Liu, T.~Lu, Z.~Li, and P.~Luo, ``Ddp: Diffusion model for dense visual prediction,'' in {\em Proceedings of the IEEE International Conference on Computer Vision}, pp.~21741--21752, 2023.

\bibitem{wei2024diff}
T.~Wei, Y.~Lin, S.~Guo, Y.~Lin, Y.~Huang, C.~Xiang, Y.~Bai, M.~Ya, and H.~Wan, ``Diff-rntraj: A structure-aware diffusion model for road network-constrained trajectory generation,'' {\em arXiv preprint arXiv:2402.07369}, 2024.

\bibitem{zou2024diffbev}
J.~Zou, K.~Tian, Z.~Zhu, Y.~Ye, and X.~Wang, ``Diffbev: Conditional diffusion model for bird’s eye view perception,'' in {\em Proceedings of the AAAI Conference on Artificial Intelligence}, vol.~38, pp.~7846--7854, 2024.

\bibitem{li2023drivingdiffusion}
X.~Li, Y.~Zhang, and X.~Ye, ``Drivingdiffusion: Layout-guided multi-view driving scene video generation with latent diffusion model,'' {\em arXiv preprint arXiv:2310.07771}, 2023.

\bibitem{zhu2023difftraj}
Y.~Zhu, Y.~Ye, S.~Zhang, X.~Zhao, and J.~Yu, ``Difftraj: Generating gps trajectory with diffusion probabilistic model,'' {\em Advances in Neural Information Processing Systems}, vol.~36, pp.~65168--65188, 2023.

\bibitem{wu2023diffumask}
W.~Wu, Y.~Zhao, M.~Z. Shou, H.~Zhou, and C.~Shen, ``Diffumask: Synthesizing images with pixel-level annotations for semantic segmentation using diffusion models,'' in {\em Proceedings of the IEEE International Conference on Computer Vision}, pp.~1206--1217, 2023.

\bibitem{zhang2023chattraffc}
C.~Zhang, Y.~Zhang, Q.~Shao, B.~Li, Y.~Lv, X.~Piao, and B.~Yin, ``Chattraffc: Text-to-traffic generation via diffusion model,'' {\em arXiv preprint arXiv:2311.16203}, 2023.

\bibitem{pronovost2023scenario}
E.~Pronovost, M.~R. Ganesina, N.~Hendy, Z.~Wang, A.~Morales, K.~Wang, and N.~Roy, ``Scenario diffusion: Controllable driving scenario generation with diffusion,'' {\em Advances in Neural Information Processing Systems}, vol.~36, pp.~68873--68894, 2023.

\bibitem{niedoba2024diffusion}
M.~Niedoba, J.~Lavington, Y.~Liu, V.~Lioutas, J.~Sefas, X.~Liang, D.~Green, S.~Dabiri, B.~Zwartsenberg, A.~Scibior, {\em et~al.}, ``A diffusion-model of joint interactive navigation,'' {\em Advances in Neural Information Processing Systems}, vol.~36, 2024.

\bibitem{esser2021taming}
P.~Esser, R.~Rombach, and B.~Ommer, ``Taming transformers for high-resolution image synthesis,'' in {\em Proceedings of the IEEE conference on computer vision and pattern recognition}, pp.~12873--12883, 2021.

\bibitem{blattmann2023align}
A.~Blattmann, R.~Rombach, H.~Ling, T.~Dockhorn, S.~W. Kim, S.~Fidler, and K.~Kreis, ``Align your latents: High-resolution video synthesis with latent diffusion models,'' in {\em Proceedings of the IEEE Conference on Computer Vision and Pattern Recognition}, pp.~22563--22575, 2023.

\bibitem{balasubramanian2023scenediffusion}
L.~Balasubramanian, J.~Wurst, R.~Egolf, M.~Botsch, W.~Utschick, and K.~Deng, ``Scenediffusion: Conditioned latent diffusion models for traffic scene prediction,'' in {\em 2023 IEEE 26th International Conference on Intelligent Transportation Systems}, pp.~3914--3921, 2023.

\bibitem{wang2024dragtraffic}
S.~Wang, G.~Sun, F.~Ma, T.~Hu, Y.~Song, L.~Zhu, and M.~Liu, ``Dragtraffic: A non-expert interactive and point-based controllable traffic scene generation framework,'' {\em arXiv preprint arXiv:2404.12624}, 2024.

\bibitem{wang2024driving}
Y.~Wang, J.~He, L.~Fan, H.~Li, Y.~Chen, and Z.~Zhang, ``Driving into the future: Multiview visual forecasting and planning with world model for autonomous driving,'' in {\em Proceedings of the IEEE Conference on Computer Vision and Pattern Recognition}, pp.~14749--14759, 2024.

\bibitem{yang2024generalized}
J.~Yang, S.~Gao, Y.~Qiu, L.~Chen, T.~Li, B.~Dai, K.~Chitta, P.~Wu, J.~Zeng, P.~Luo, {\em et~al.}, ``Generalized predictive model for autonomous driving,'' in {\em Proceedings of the IEEE Conference on Computer Vision and Pattern Recognition}, pp.~14662--14672, 2024.

\bibitem{wang2023drivedreamer}
X.~Wang, Z.~Zhu, G.~Huang, X.~Chen, and J.~Lu, ``Drivedreamer: Towards real-world-driven world models for autonomous driving,'' {\em arXiv preprint arXiv:2309.09777}, 2023.

\bibitem{lu2023wovogen}
J.~Lu, Z.~Huang, J.~Zhang, Z.~Yang, and L.~Zhang, ``Wovogen: World volume-aware diffusion for controllable multi-camera driving scene generation,'' {\em arXiv preprint arXiv:2312.02934}, 2023.

\bibitem{ran2024towards}
H.~Ran, V.~Guizilini, and Y.~Wang, ``Towards realistic scene generation with lidar diffusion models,'' in {\em Proceedings of the IEEE Conference on Computer Vision and Pattern Recognition}, pp.~14738--14748, 2024.

\bibitem{zheng2023diffuflow}
Y.~Zheng, L.~Zhong, S.~Wang, Y.~Yang, W.~Gu, J.~Zhang, and J.~Wang, ``Diffuflow: Robust fine-grained urban flow inference with denoising diffusion model,'' in {\em Proceedings of the 32nd ACM International Conference on Information and Knowledge Management}, pp.~3505--3513, 2023.

\bibitem{elman1990finding}
J.~L. Elman, ``Finding structure in time,'' {\em Cognitive Science}, vol.~14, no.~2, pp.~179--211, 1990.

\bibitem{hochreiter1997long}
S.~Hochreiter, ``Long short-term memory,'' {\em Neural Computation MIT-Press}, 1997.

\bibitem{scarselli2008graph}
F.~Scarselli, M.~Gori, A.~C. Tsoi, M.~Hagenbuchner, and G.~Monfardini, ``The graph neural network model,'' {\em IEEE Transactions on Neural Networks}, vol.~20, no.~1, pp.~61--80, 2008.

\bibitem{kipf2016semi}
T.~N. Kipf and M.~Welling, ``Semi-supervised classification with graph convolutional networks,'' {\em arXiv preprint arXiv:1609.02907}, 2016.

\bibitem{goodfellow2014generative}
I.~Goodfellow, J.~Pouget-Abadie, M.~Mirza, B.~Xu, D.~Warde-Farley, S.~Ozair, A.~Courville, and Y.~Bengio, ``Generative adversarial nets,'' {\em Advances in Neural Information Processing Systems}, vol.~27, 2014.

\bibitem{creswell2018generative}
A.~Creswell, T.~White, V.~Dumoulin, K.~Arulkumaran, B.~Sengupta, and A.~A. Bharath, ``Generative adversarial networks: An overview,'' {\em IEEE Signal Processing Magazine}, vol.~35, no.~1, pp.~53--65, 2018.

\bibitem{kingma2013auto}
D.~P. Kingma, ``Auto-encoding variational bayes,'' {\em arXiv preprint arXiv:1312.6114}, 2013.

\bibitem{rezende2014stochastic}
D.~J. Rezende, S.~Mohamed, and D.~Wierstra, ``Stochastic backpropagation and approximate inference in deep generative models,'' in {\em International Conference on Machine Learning}, pp.~1278--1286, 2014.

\bibitem{zhao2023autonomous}
J.~Zhao, W.~Zhao, B.~Deng, Z.~Wang, F.~Zhang, W.~Zheng, W.~Cao, J.~Nan, Y.~Lian, and A.~F. Burke, ``Autonomous driving system: A comprehensive survey,'' {\em Expert Systems with Applications}, p.~122836, 2023.

\bibitem{chen2023end}
L.~Chen, P.~Wu, K.~Chitta, B.~Jaeger, A.~Geiger, and H.~Li, ``End-to-end autonomous driving: Challenges and frontiers,'' {\em arXiv preprint arXiv:2306.16927}, 2023.

\bibitem{teng2023motion}
S.~Teng, X.~Hu, P.~Deng, B.~Li, Y.~Li, Y.~Ai, D.~Yang, L.~Li, Z.~Xuanyuan, F.~Zhu, {\em et~al.}, ``Motion planning for autonomous driving: The state of the art and future perspectives,'' {\em IEEE Transactions on Intelligent Vehicles}, vol.~8, no.~6, pp.~3692--3711, 2023.

\bibitem{grigorescu2020survey}
S.~Grigorescu, B.~Trasnea, T.~Cocias, and G.~Macesanu, ``A survey of deep learning techniques for autonomous driving,'' {\em Journal of field robotics}, vol.~37, no.~3, pp.~362--386, 2020.

\bibitem{yurtsever2020survey}
E.~Yurtsever, J.~Lambert, A.~Carballo, and K.~Takeda, ``A survey of autonomous driving: Common practices and emerging technologies,'' {\em IEEE access}, vol.~8, pp.~58443--58469, 2020.

\bibitem{baranchuk2021label}
D.~Baranchuk, I.~Rubachev, A.~Voynov, V.~Khrulkov, and A.~Babenko, ``Label-efficient semantic segmentation with diffusion models,'' {\em arXiv preprint arXiv:2112.03126}, 2021.

\bibitem{le2024diffuser}
D.-T. Le, H.~Shi, J.~Cai, and H.~Rezatofighi, ``Diffuser: Diffusion model for robust multi-sensor fusion in 3d object detection and bev segmentation,'' {\em arXiv preprint arXiv:2404.04629}, 2024.

\bibitem{li2024light}
J.~Li, B.~Li, Z.~Tu, X.~Liu, Q.~Guo, F.~Juefei-Xu, R.~Xu, and H.~Yu, ``Light the night: A multi-condition diffusion framework for unpaired low-light enhancement in autonomous driving,'' in {\em Proceedings of the IEEE Conference on Computer Vision and Pattern Recognition}, pp.~15205--15215, 2024.

\bibitem{nachkov2023diffusion}
A.~Nachkov, M.~Danelljan, D.~P. Paudel, and L.~Van~Gool, ``Diffusion-based particle-detr for bev perception,'' {\em arXiv preprint arXiv:2312.11578}, 2023.

\bibitem{chen2023diffusiondet}
S.~Chen, P.~Sun, Y.~Song, and P.~Luo, ``Diffusiondet: Diffusion model for object detection,'' in {\em Proceedings of the IEEE international conference on computer vision}, pp.~19830--19843, 2023.

\bibitem{wang2024detdiffusion}
Y.~Wang, R.~Gao, K.~Chen, K.~Zhou, Y.~Cai, L.~Hong, Z.~Li, L.~Jiang, D.-Y. Yeung, Q.~Xu, {\em et~al.}, ``Detdiffusion: Synergizing generative and perceptive models for enhanced data generation and perception,'' in {\em Proceedings of the IEEE Conference on Computer Vision and Pattern Recognition}, pp.~7246--7255, 2024.

\bibitem{zhao2023unleashing}
W.~Zhao, Y.~Rao, Z.~Liu, B.~Liu, J.~Zhou, and J.~Lu, ``Unleashing text-to-image diffusion models for visual perception,'' in {\em Proceedings of the IEEE International Conference on Computer Vision}, pp.~5729--5739, 2023.

\bibitem{luo2021multiple}
W.~Luo, J.~Xing, A.~Milan, X.~Zhang, W.~Liu, and T.-K. Kim, ``Multiple object tracking: A literature review,'' {\em Artificial intelligence}, vol.~293, p.~103448, 2021.

\bibitem{chen2024delving}
S.~Chen, E.~Yu, J.~Li, and W.~Tao, ``Delving into the trajectory long-tail distribution for muti-object tracking,'' in {\em Proceedings of the IEEE Conference on Computer Vision and Pattern Recognition}, pp.~19341--19351, 2024.

\bibitem{luo2024diffusiontrack}
R.~Luo, Z.~Song, L.~Ma, J.~Wei, W.~Yang, and M.~Yang, ``Diffusiontrack: Diffusion model for multi-object tracking,'' in {\em Proceedings of the AAAI Conference on Artificial Intelligence}, vol.~38, pp.~3991--3999, 2024.

\bibitem{xie2024diffusiontrack}
F.~Xie, Z.~Wang, and C.~Ma, ``Diffusiontrack: Point set diffusion model for visual object tracking,'' in {\em Proceedings of the IEEE Conference on Computer Vision and Pattern Recognition}, pp.~19113--19124, 2024.

\bibitem{wu2023datasetdm}
W.~Wu, Y.~Zhao, H.~Chen, Y.~Gu, R.~Zhao, Y.~He, H.~Zhou, M.~Z. Shou, and C.~Shen, ``Datasetdm: Synthesizing data with perception annotations using diffusion models,'' {\em Advances in Neural Information Processing Systems}, vol.~36, pp.~54683--54695, 2023.

\bibitem{luo2021diffusion}
S.~Luo and W.~Hu, ``Diffusion probabilistic models for 3d point cloud generation,'' in {\em Proceedings of the IEEE conference on computer vision and pattern recognition}, pp.~2837--2845, 2021.

\bibitem{sun2022pointdp}
J.~Sun, W.~Nie, Z.~Yu, Z.~M. Mao, and C.~Xiao, ``Pointdp: Diffusion-driven purification against adversarial attacks on 3d point cloud recognition,'' {\em arXiv preprint arXiv:2208.09801}, 2022.

\bibitem{huang2022survey}
Y.~Huang, J.~Du, Z.~Yang, Z.~Zhou, L.~Zhang, and H.~Chen, ``A survey on trajectory-prediction methods for autonomous driving,'' {\em IEEE Transactions on Intelligent Vehicles}, vol.~7, no.~3, pp.~652--674, 2022.

\bibitem{huang2023multimodal}
R.~Huang, H.~Xue, M.~Pagnucco, F.~Salim, and Y.~Song, ``Multimodal trajectory prediction: A survey,'' {\em arXiv preprint arXiv:2302.10463}, 2023.

\bibitem{mao2023leapfrog}
W.~Mao, C.~Xu, Q.~Zhu, S.~Chen, and Y.~Wang, ``Leapfrog diffusion model for stochastic trajectory prediction,'' in {\em Proceedings of the IEEE conference on computer vision and pattern recognition}, pp.~5517--5526, 2023.

\bibitem{bae2024singulartrajectory}
I.~Bae, Y.-J. Park, and H.-G. Jeon, ``Singulartrajectory: Universal trajectory predictor using diffusion model,'' in {\em Proceedings of the IEEE Conference on Computer Vision and Pattern Recognition}, pp.~17890--17901, 2024.

\bibitem{liu2024intention}
C.~Liu, S.~He, H.~Liu, and J.~Chen, ``Intention-aware denoising diffusion model for trajectory prediction,'' {\em arXiv preprint arXiv:2403.09190}, 2024.

\bibitem{lv2024learning}
K.~Lv, L.~Yuan, and X.~Ni, ``Learning autoencoder diffusion models of pedestrian group relationships for multimodal trajectory prediction,'' {\em IEEE Transactions on Instrumentation and Measurement}, 2024.

\bibitem{sun2022human}
J.~Sun, Y.~Li, L.~Chai, H.-S. Fang, Y.-L. Li, and C.~Lu, ``Human trajectory prediction with momentary observation,'' in {\em Proceedings of the IEEE Conference on Computer Vision and Pattern Recognition}, pp.~6467--6476, 2022.

\bibitem{li2024bcdiff}
R.~Li, C.~Li, D.~Ren, G.~Chen, Y.~Yuan, and G.~Wang, ``Bcdiff: Bidirectional consistent diffusion for instantaneous trajectory prediction,'' {\em Advances in Neural Information Processing Systems}, vol.~36, 2024.

\bibitem{westny2024diffusion}
T.~Westny, B.~Olofsson, and E.~Frisk, ``Diffusion-based environment-aware trajectory prediction,'' {\em arXiv preprint arXiv:2403.11643}, 2024.

\bibitem{yao2023graph}
Y.~Yao, Y.~Liu, X.~Dai, S.~Chen, and Y.~Lv, ``A graph-based scene encoder for vehicle trajectory prediction using the diffusion model,'' in {\em 2023 International Annual Conference on Complex Systems and Intelligent Science}, pp.~981--986, 2023.

\bibitem{badue2021self}
C.~Badue, R.~Guidolini, R.~V. Carneiro, P.~Azevedo, V.~B. Cardoso, A.~Forechi, L.~Jesus, R.~Berriel, T.~M. Paixao, F.~Mutz, {\em et~al.}, ``Self-driving cars: A survey,'' {\em Expert systems with applications}, vol.~165, p.~113816, 2021.

\bibitem{carvalho2023motion}
J.~Carvalho, A.~T. Le, M.~Baierl, D.~Koert, and J.~Peters, ``Motion planning diffusion: Learning and planning of robot motions with diffusion models,'' in {\em 2023 IEEE International Conference on Intelligent Robots and Systems}, pp.~1916--1923, 2023.

\bibitem{ajay2022conditional}
A.~Ajay, Y.~Du, A.~Gupta, J.~Tenenbaum, T.~Jaakkola, and P.~Agrawal, ``Is conditional generative modeling all you need for decision-making?,'' {\em arXiv preprint arXiv:2211.15657}, 2022.

\bibitem{yang2024diffusion}
B.~Yang, H.~Su, N.~Gkanatsios, T.-W. Ke, A.~Jain, J.~Schneider, and K.~Fragkiadaki, ``Diffusion-es: Gradient-free planning with diffusion for autonomous and instruction-guided driving,'' in {\em Proceedings of the IEEE Conference on Computer Vision and Pattern Recognition}, pp.~15342--15353, 2024.

\bibitem{chen2024human}
K.~Chen, Y.~Luo, M.~Zhu, and H.~Yang, ``Human-like interactive lane-change modeling based on reward-guided diffusive predictor and planner,'' {\em IEEE Transactions on Intelligent Transportation Systems}, 2024.

\bibitem{chen2024dynamic}
D.~Chen, R.~Zhong, K.~Chen, Z.~Shang, M.~Zhu, and E.~Chung, ``Dynamic high-order control barrier functions with diffuser for safety-critical trajectory planning at signal-free intersections,'' {\em arXiv preprint arXiv:2412.00162}, 2024.

\bibitem{kiran2021deep}
B.~R. Kiran, I.~Sobh, V.~Talpaert, P.~Mannion, A.~A. Al~Sallab, S.~Yogamani, and P.~P{\'e}rez, ``Deep reinforcement learning for autonomous driving: A survey,'' {\em IEEE Transactions on Intelligent Transportation Systems}, vol.~23, no.~6, pp.~4909--4926, 2021.

\bibitem{aradi2020survey}
S.~Aradi, ``Survey of deep reinforcement learning for motion planning of autonomous vehicles,'' {\em IEEE Transactions on Intelligent Transportation Systems}, vol.~23, no.~2, pp.~740--759, 2020.

\bibitem{zhu2023diffusion}
Z.~Zhu, H.~Zhao, H.~He, Y.~Zhong, S.~Zhang, Y.~Yu, and W.~Zhang, ``Diffusion models for reinforcement learning: A survey,'' {\em arXiv preprint arXiv:2311.01223}, 2023.

\bibitem{wang2022diffusion}
Z.~Wang, J.~J. Hunt, and M.~Zhou, ``Diffusion policies as an expressive policy class for offline reinforcement learning,'' {\em arXiv preprint arXiv:2208.06193}, 2022.

\bibitem{liu2024ddm}
J.~Liu, P.~Hang, X.~Zhao, J.~Wang, and J.~Sun, ``Ddm-lag: A diffusion-based decision-making model for autonomous vehicles with lagrangian safety enhancement,'' {\em arXiv preprint arXiv:2401.03629}, 2024.

\bibitem{nguyen2021overview}
J.~Nguyen, S.~T. Powers, N.~Urquhart, T.~Farrenkopf, and M.~Guckert, ``An overview of agent-based traffic simulators,'' {\em Transportation research interdisciplinary perspectives}, vol.~12, p.~100486, 2021.

\bibitem{chen2024data}
D.~Chen, M.~Zhu, H.~Yang, X.~Wang, and Y.~Wang, ``Data-driven traffic simulation: A comprehensive review,'' {\em IEEE Transactions on Intelligent Vehicles}, 2024.

\bibitem{ding2023survey}
W.~Ding, C.~Xu, M.~Arief, H.~Lin, B.~Li, and D.~Zhao, ``A survey on safety-critical driving scenario generation—a methodological perspective,'' {\em IEEE Transactions on Intelligent Transportation Systems}, vol.~24, no.~7, pp.~6971--6988, 2023.

\bibitem{lopez2018microscopic}
P.~A. Lopez, M.~Behrisch, L.~Bieker-Walz, J.~Erdmann, Y.-P. Fl{\"o}tter{\"o}d, R.~Hilbrich, L.~L{\"u}cken, J.~Rummel, P.~Wagner, and E.~Wie{\ss}ner, ``Microscopic traffic simulation using sumo,'' in {\em 2018 21st international conference on intelligent transportation systems}, pp.~2575--2582, 2018.

\bibitem{xie2024advdiffuser}
Y.~Xie, X.~Guo, C.~Wang, K.~Liu, and L.~Chen, ``Advdiffuser: Generating adversarial safety-critical driving scenarios via guided diffusion,'' in {\em 2024 IEEE International Conference on Intelligent Robots and Systems}, pp.~9983--9989, 2024.

\bibitem{zhong2023language}
Z.~Zhong, D.~Rempe, Y.~Chen, B.~Ivanovic, Y.~Cao, D.~Xu, M.~Pavone, and B.~Ray, ``Language-guided traffic simulation via scene-level diffusion,'' in {\em Conference on Robot Learning}, pp.~144--177, PMLR, 2023.

\bibitem{huang2024versatile}
Z.~Huang, Z.~Zhang, A.~Vaidya, Y.~Chen, C.~Lv, and J.~F. Fisac, ``Versatile scene-consistent traffic scenario generation as optimization with diffusion,'' {\em arXiv preprint arXiv:2404.02524}, 2024.

\bibitem{yang2024wcdt}
C.~Yang, A.~X. Tian, D.~Chen, T.~Shi, and A.~Heydarian, ``Wcdt: World-centric diffusion transformer for traffic scene generation,'' {\em arXiv preprint arXiv:2404.02082}, 2024.

\bibitem{rempe2023trace}
D.~Rempe, Z.~Luo, X.~Bin~Peng, Y.~Yuan, K.~Kitani, K.~Kreis, S.~Fidler, and O.~Litany, ``Trace and pace: Controllable pedestrian animation via guided trajectory diffusion,'' in {\em Proceedings of the IEEE Conference on Computer Vision and Pattern Recognition}, pp.~13756--13766, 2023.

\bibitem{riedmaier2020survey}
S.~Riedmaier, T.~Ponn, D.~Ludwig, B.~Schick, and F.~Diermeyer, ``Survey on scenario-based safety assessment of automated vehicles,'' {\em IEEE access}, vol.~8, pp.~87456--87477, 2020.

\bibitem{zhu2024sora}
Z.~Zhu, X.~Wang, W.~Zhao, C.~Min, N.~Deng, M.~Dou, Y.~Wang, B.~Shi, K.~Wang, C.~Zhang, {\em et~al.}, ``Is sora a world simulator? a comprehensive survey on general world models and beyond,'' {\em arXiv preprint arXiv:2405.03520}, 2024.

\bibitem{zhang2023adding}
L.~Zhang, A.~Rao, and M.~Agrawala, ``Adding conditional control to text-to-image diffusion models,'' in {\em Proceedings of the IEEE International Conference on Computer Vision}, pp.~3836--3847, 2023.

\bibitem{ho2022video}
J.~Ho, T.~Salimans, A.~Gritsenko, W.~Chan, M.~Norouzi, and D.~J. Fleet, ``Video diffusion models,'' {\em Advances in Neural Information Processing Systems}, vol.~35, pp.~8633--8646, 2022.

\bibitem{ho2022imagen}
J.~Ho, W.~Chan, C.~Saharia, J.~Whang, R.~Gao, A.~Gritsenko, D.~P. Kingma, B.~Poole, M.~Norouzi, D.~J. Fleet, {\em et~al.}, ``Imagen video: High definition video generation with diffusion models,'' {\em arXiv preprint arXiv:2210.02303}, 2022.

\bibitem{ha2018recurrent}
D.~Ha and J.~Schmidhuber, ``Recurrent world models facilitate policy evolution,'' {\em Advances in Neural Information Processing Systems}, vol.~31, 2018.

\bibitem{harvey2022flexible}
W.~Harvey, S.~Naderiparizi, V.~Masrani, C.~Weilbach, and F.~Wood, ``Flexible diffusion modeling of long videos,'' {\em Advances in Neural Information Processing Systems}, vol.~35, pp.~27953--27965, 2022.

\bibitem{zhao2024drivedreamer}
G.~Zhao, X.~Wang, Z.~Zhu, X.~Chen, G.~Huang, X.~Bao, and X.~Wang, ``Drivedreamer-2: Llm-enhanced world models for diverse driving video generation,'' {\em arXiv preprint arXiv:2403.06845}, 2024.

\bibitem{mao2023gpt}
J.~Mao, Y.~Qian, H.~Zhao, and Y.~Wang, ``Gpt-driver: Learning to drive with gpt,'' {\em arXiv preprint arXiv:2310.01415}, 2023.

\bibitem{peng2024lc}
M.~Peng, X.~Guo, X.~Chen, M.~Zhu, K.~Chen, X.~Wang, Y.~Wang, {\em et~al.}, ``Lc-llm: Explainable lane-change intention and trajectory predictions with large language models,'' {\em arXiv preprint arXiv:2403.18344}, 2024.

\bibitem{zhang2023learning}
L.~Zhang, Y.~Xiong, Z.~Yang, S.~Casas, R.~Hu, and R.~Urtasun, ``Learning unsupervised world models for autonomous driving via discrete diffusion,'' {\em arXiv preprint arXiv:2311.01017}, 2023.

\bibitem{zyrianov2024lidardm}
V.~Zyrianov, H.~Che, Z.~Liu, and S.~Wang, ``Lidardm: Generative lidar simulation in a generated world,'' {\em arXiv preprint arXiv:2404.02903}, 2024.

\bibitem{radford2021learning}
A.~Radford, J.~W. Kim, C.~Hallacy, A.~Ramesh, G.~Goh, S.~Agarwal, G.~Sastry, A.~Askell, P.~Mishkin, J.~Clark, {\em et~al.}, ``Learning transferable visual models from natural language supervision,'' in {\em International Conference on Machine Learning}, pp.~8748--8763, PMLR, 2021.

\bibitem{van2017neural}
A.~Van Den~Oord, O.~Vinyals, {\em et~al.}, ``Neural discrete representation learning,'' {\em Advances in Neural Information Processing Systems}, vol.~30, 2017.

\bibitem{chang2022maskgit}
H.~Chang, H.~Zhang, L.~Jiang, C.~Liu, and W.~T. Freeman, ``Maskgit: Masked generative image transformer,'' in {\em Proceedings of the IEEE Conference on Computer Vision and Pattern Recognition}, pp.~11315--11325, 2022.

\bibitem{austin2021structured}
J.~Austin, D.~D. Johnson, J.~Ho, D.~Tarlow, and R.~Van Den~Berg, ``Structured denoising diffusion models in discrete state-spaces,'' {\em Advances in Neural Information Processing Systems}, vol.~34, pp.~17981--17993, 2021.

\bibitem{rasul2021autoregressive}
K.~Rasul, C.~Seward, I.~Schuster, and R.~Vollgraf, ``Autoregressive denoising diffusion models for multivariate probabilistic time series forecasting,'' in {\em International Conference on Machine Learning}, pp.~8857--8868, 2021.

\bibitem{yuan2023spatio}
Y.~Yuan, J.~Ding, C.~Shao, D.~Jin, and Y.~Li, ``Spatio-temporal diffusion point processes,'' in {\em Proceedings of the 29th ACM Conference on Knowledge Discovery and Data Mining}, pp.~3173--3184, 2023.

\bibitem{zhou2023towards}
Z.~Zhou, J.~Ding, Y.~Liu, D.~Jin, and Y.~Li, ``Towards generative modeling of urban flow through knowledge-enhanced denoising diffusion,'' in {\em Proceedings of the 31st ACM International Conference on Advances in Geographic Information Systems}, pp.~1--12, 2023.

\bibitem{devlin2018bert}
J.~Devlin, M.-W. Chang, K.~Lee, and K.~Toutanova, ``Bert: Pre-training of deep bidirectional transformers for language understanding,'' {\em arXiv preprint arXiv:1810.04805}, 2018.

\bibitem{rong2023complexity}
C.~Rong, J.~Ding, Z.~Liu, and Y.~Li, ``Complexity-aware large scale origin-destination network generation via diffusion model,'' {\em arXiv preprint arXiv:2306.04873}, 2023.

\bibitem{jiang2022graph}
W.~Jiang and J.~Luo, ``Graph neural network for traffic forecasting: A survey,'' {\em Expert Systems with Applications}, vol.~207, p.~117921, 2022.

\bibitem{zheng2024recovering}
Z.~Zheng, Z.~Wang, Z.~Hu, Z.~Wan, and W.~Ma, ``Recovering traffic data from the corrupted noise: A doubly physics-regularized denoising diffusion model,'' {\em Transportation Research Part C: Emerging Technologies}, vol.~160, p.~104513, 2024.

\bibitem{lin2024specstg}
L.~Lin, D.~Shi, A.~Han, and J.~Gao, ``Specstg: A fast spectral diffusion framework for probabilistic spatio-temporal traffic forecasting,'' {\em arXiv preprint arXiv:2401.08119}, 2024.

\bibitem{xu2023diffusion}
X.~Xu, Y.~Wei, P.~Wang, X.~Luo, F.~Zhou, and G.~Trajcevski, ``Diffusion probabilistic modeling for fine-grained urban traffic flow inference with relaxed structural constraint,'' in {\em IEEE International Conference on Acoustics, Speech and Signal Processing}, pp.~1--5, 2023.

\bibitem{lablack2023long}
M.~Lablack, S.~Yu, S.~Xu, and Y.~Shen, ``Long-sequence model for traffic forecasting in suboptimal situation,'' in {\em Proceedings of the 18th Workshop on Mobility in the Evolving Internet Architecture}, pp.~25--30, 2023.

\bibitem{chi2023difforecast}
P.~Chi and X.~Ma, ``Difforecast: Image generation based highway traffic forecasting with diffusion model,'' in {\em 2023 IEEE International Conference on Big Data}, pp.~608--615, 2023.

\bibitem{mori2015review}
U.~Mori, A.~Mendiburu, M.~{\'A}lvarez, and J.~A. Lozano, ``A review of travel time estimation and forecasting for advanced traveller information systems,'' {\em Transportmetrica A: Transport Science}, vol.~11, no.~2, pp.~119--157, 2015.

\bibitem{lin2023origin}
Y.~Lin, H.~Wan, J.~Hu, S.~Guo, B.~Yang, Y.~Lin, and C.~S. Jensen, ``Origin-destination travel time oracle for map-based services,'' {\em Proceedings of the ACM on Management of Data}, vol.~1, no.~3, pp.~1--27, 2023.

\bibitem{goniewicz2016road}
K.~Goniewicz, M.~Goniewicz, W.~Paw{\l}owski, and P.~Fiedor, ``Road accident rates: strategies and programmes for improving road traffic safety,'' {\em European Journal of Trauma and Emergency Surgery}, vol.~42, pp.~433--438, 2016.

\bibitem{santhosh2020anomaly}
K.~K. Santhosh, D.~P. Dogra, and P.~P. Roy, ``Anomaly detection in road traffic using visual surveillance: A survey,'' {\em ACM Computing Surveys}, vol.~53, no.~6, pp.~1--26, 2020.

\bibitem{kong2024mobile}
X.~Kong, J.~Wang, Z.~Hu, Y.~He, X.~Zhao, and G.~Shen, ``Mobile trajectory anomaly detection: Taxonomy, methodology, challenges, and directions,'' {\em IEEE Internet of Things Journal}, 2024.

\bibitem{li2024difftad}
C.~Li, G.~Feng, Y.~Li, R.~Liu, Q.~Miao, and L.~Chang, ``Difftad: Denoising diffusion probabilistic models for vehicle trajectory anomaly detection,'' {\em Knowledge-Based Systems}, vol.~286, p.~111387, 2024.

\bibitem{yan2023feature}
C.~Yan, S.~Zhang, Y.~Liu, G.~Pang, and W.~Wang, ``Feature prediction diffusion model for video anomaly detection,'' in {\em Proceedings of the IEEE International Conference on Computer Vision}, pp.~5527--5537, 2023.

\bibitem{fang2023vision}
J.~Fang, J.~Qiao, J.~Xue, and Z.~Li, ``Vision-based traffic accident detection and anticipation: A survey,'' {\em IEEE Transactions on Circuits and Systems for Video Technology}, 2023.

\bibitem{fang2024abductive}
J.~Fang, L.-l. Li, J.~Zhou, J.~Xiao, H.~Yu, C.~Lv, J.~Xue, and T.-S. Chua, ``Abductive ego-view accident video understanding for safe driving perception,'' in {\em Proceedings of the IEEE Conference on Computer Vision and Pattern Recognition}, pp.~22030--22040, 2024.

\bibitem{zheng2023minigpt}
K.~Zheng, X.~He, and X.~E. Wang, ``Minigpt-5: Interleaved vision-and-language generation via generative vokens,'' {\em arXiv preprint arXiv:2310.02239}, 2023.

\bibitem{zhao2023making}
X.~Zhao, B.~Liu, Q.~Liu, G.~Shi, and X.-M. Wu, ``Making multimodal generation easier: When diffusion models meet llms,'' {\em arXiv preprint arXiv:2310.08949}, 2023.

\bibitem{zhang2024chatscene}
J.~Zhang, C.~Xu, and B.~Li, ``Chatscene: Knowledge-enabled safety-critical scenario generation for autonomous vehicles,'' in {\em Proceedings of the IEEE Conference on Computer Vision and Pattern Recognition}, pp.~15459--15469, 2024.

\bibitem{peebles2023scalable}
W.~Peebles and S.~Xie, ``Scalable diffusion models with transformers,'' in {\em Proceedings of the IEEE International Conference on Computer Vision}, pp.~4195--4205, 2023.

\bibitem{bao2023all}
F.~Bao, S.~Nie, K.~Xue, Y.~Cao, C.~Li, H.~Su, and J.~Zhu, ``All are worth words: A vit backbone for diffusion models,'' in {\em Proceedings of the IEEE Conference on Computer Cision and Pattern Recognition}, pp.~22669--22679, 2023.

\bibitem{esser2024scaling}
P.~Esser, S.~Kulal, A.~Blattmann, R.~Entezari, J.~M{\"u}ller, H.~Saini, Y.~Levi, D.~Lorenz, A.~Sauer, F.~Boesel, {\em et~al.}, ``Scaling rectified flow transformers for high-resolution image synthesis,'' in {\em Forty-first International Conference on Machine Learning}, 2024.

\bibitem{mou2024t2i}
C.~Mou, X.~Wang, L.~Xie, Y.~Wu, J.~Zhang, Z.~Qi, and Y.~Shan, ``T2i-adapter: Learning adapters to dig out more controllable ability for text-to-image diffusion models,'' in {\em Proceedings of the AAAI Conference on Artificial Intelligence}, vol.~38, pp.~4296--4304, 2024.

\bibitem{song2020denoising}
J.~Song, C.~Meng, and S.~Ermon, ``Denoising diffusion implicit models,'' {\em arXiv preprint arXiv:2010.02502}, 2020.

\bibitem{salimans2022progressive}
T.~Salimans and J.~Ho, ``Progressive distillation for fast sampling of diffusion models,'' {\em arXiv preprint arXiv:2202.00512}, 2022.

\bibitem{song2023consistency}
Y.~Song, P.~Dhariwal, M.~Chen, and I.~Sutskever, ``Consistency models,'' {\em arXiv preprint arXiv:2303.01469}, 2023.

\bibitem{xue2024accelerating}
S.~Xue, Z.~Liu, F.~Chen, S.~Zhang, T.~Hu, E.~Xie, and Z.~Li, ``Accelerating diffusion sampling with optimized time steps,'' in {\em Proceedings of the IEEE Conference on Computer Vision and Pattern Recognition}, pp.~8292--8301, 2024.

\bibitem{karras2022elucidating}
T.~Karras, M.~Aittala, T.~Aila, and S.~Laine, ``Elucidating the design space of diffusion-based generative models,'' {\em Advances in Neural Information Processing Systems}, vol.~35, pp.~26565--26577, 2022.

\bibitem{han2022card}
X.~Han, H.~Zheng, and M.~Zhou, ``Card: Classification and regression diffusion models,'' {\em Advances in Neural Information Processing Systems}, vol.~35, pp.~18100--18115, 2022.

\end{thebibliography}

\newpage

\onecolumn
\newpage
\appendix
\section{Supplementary Materials}
\definecolor{LightRed}{rgb}{1,0.92,0.92}
\definecolor{LightOrange}{rgb}{1,0.95,0.88}
\definecolor{LightYellow}{rgb}{1.0,1.0,0.84}
\definecolor{LightGreen}{rgb}{0.9,1.0,0.88}
\definecolor{LightCyan}{rgb}{0.9,1,1}
\definecolor{LightBlue}{rgb}{0.85, 0.94, 1.0}
\definecolor{LightBlue2}{rgb}{0.88, 0.92, 1.0}
\definecolor{LightIndigo}{rgb}{0.92,0.88, 0.98}
\definecolor{LightMagenta}{rgb}{0.96,0.86, 0.96}
\definecolor{DirtyWhite}{rgb}{0.96,0.96,0.96}

\begin{table*}[!th]
\setlength\tabcolsep{2pt}
\caption{Applications of diffusion models in intelligent transportation systems. Three key criteria are considered to classify existing models: the task, the denoising condition, and the architecture. Additionally, the datasets and open source are provided. The following abbreviations are used to denote the architectures: DDPM (Denoising Diffusion Probabilistic Model) \cite{ho2020denoising}, DDIM (Denoising Diffusion Implicit Model) \cite{song2020denoising}, ADM (Ablated Diffusion Model) \cite{dhariwal2021diffusion}, LDM (Latent Diffusion Model) \cite{rombach2022high}, LED (LEapfrog Diffusion Model) \cite{mao2023leapfrog}, VLDM (Video Latent Diffusion Model) \cite{blattmann2023align}, EDM (Elucidating Diffusion Model) \cite{karras2022elucidating}, FDM (Flexible Diffusion Model) \cite{harvey2022flexible}, D3PM (Discrete Denoising Diffusion Probabilistic Model) \cite{austin2021structured}, CARD (Classification and Regression Diffusion Model) \cite{han2022card}.} 
\label{tab:tab_taxonomy}
\centering
\renewcommand{\arraystretch}{1.5}
\begin{tabular}{|p{2.2cm}|p{3.5cm}|p{3.5cm}|p{1.5cm}|p{2.4cm}| p{1.8cm}|p{2.2cm}|}
\hline
\textbf{Paper} & \textbf{Task} &\textbf{Denoising Condition} & \textbf{Architecture} & \textbf{Datasets} & \textbf{Year} & \textbf{Open Source} \\ 
\hline
\rowcolor{LightRed}
DiffusionDet \cite{chen2023diffusiondet} & 2D object detection & conditioned on image feature & DDIM & CrowdHuman \newline COCO & 2023 ICCV & \href{https://github.com/ShoufaChen/DiffusionDet}{DiffusionDet} \\
\hline
\rowcolor{LightRed}
DetDiffusion \cite{wang2024detdiffusion} & 2D object detection & conditioned on perception-aware attributes & LDM & COCO & 2024 CVPR &  ——\\
\hline
\rowcolor{LightRed}
DiffBEV \cite{zou2024diffbev} & BEV semantic segmentation \newline 3D object detection & conditioned on BEV feature & DDPM & nuScenes & 2024 AAAI & \href{https://github.com/JiayuZou2020/DiffBEV}{DiffBEV} \\
\hline
\rowcolor{LightRed}
DDP \cite{ji2023ddp} & BEV map segmentation \newline semantic segmentation \newline depth estimation & conditioned on image feature & DDIM & ADE20K \newline NYU-DepthV2 \newline KITTI et al. & 2023 ICCV & \href{https://github.com/JiYuanFeng/DDP}{DDP} \\
\hline
\rowcolor{LightRed}
VPD \cite{zhao2023unleashing} & semantic segmentation \newline image segmentation \newline depth estimation & conditioned on text & LDM & ADE20K \newline RefCOCO \newline NYU-DepthV2 & 2023 ICCV & \href{https://github.com/wl-zhao/VPD}{VPD} \\
\hline
\rowcolor{LightRed}
Chen et al. \cite{chen2024delving} & multi-object tracking & conditioned on text & LDM & MOT20 et al. & 2024 CVPR & \href{https://github.com/chen-si-jia/Trajectory-Long-tail-Distribution-for-MOT}{LtD-MOT} \\
\hline
\rowcolor{LightRed}
Luo et al. \cite{luo2024diffusiontrack} & multi-object tracking & conditioned on two adjacent raw images & DDPM & MOT20 et al. & 2024 AAAI & \href{https://github.com/RainBowLuoCS/DiffusionTrack}{DiffusionTrack} \\
\hline
\rowcolor{LightRed}
Xie et al. \cite{xie2024diffusiontrack} & object tracking & unconditional & DDIM & GOT-10k \newline LaSOT & 2024 CVPR & \href{https://github.com/VISION-SJTU/DiffusionTrack}{DiffusionTrack} \\
\hline
\rowcolor{LightRed}
Luo et al. \cite{luo2021diffusion} & 3D point cloud generation & conditioned on shape latent & DDPM & ShapeNet & 2021 CVPR & \href{https://github.com/luost26/diffusion-point-cloud}{DPC} \\
\hline
\rowcolor{LightRed}
DiffuMask \cite{wu2023diffumask} & semantic segmentation \newline perception data augmentation & conditioned on text & LDM & VOC \newline ADE20K \newline Cityscapes & 2023 ICCV & \href{https://github.com/weijiawu/DiffuMask}{DiffuMask} \\
\hline
\rowcolor{LightRed}
DatasetDM \cite{wu2023datasetdm} & perception data augmentation & conditioned on text & LDM & COCO et al. & 2023 NIPS & \href{https://github.com/showlab/DatasetDM}{DatasetDM} \\
\hline

\rowcolor{LightOrange}
MID \cite{gu2022stochastic} & human trajectory prediction & conditioned on observed trajectories & DDPM & SDD \newline ETH \newline UCY & 2022 CVPR & \href{https://github.com/gutianpei/MID}{MID} \\
\hline
\rowcolor{LightOrange}
LED \cite{mao2023leapfrog} & human trajectory prediction \newline speed up & conditioned on observed trajectories & LED & SDD et al. & 2023 CVPR & \href{https://github.com/MediaBrain-SJTU/LED}{LED} \\
\hline
\rowcolor{LightOrange}
SingularTrajectory \cite{bae2024singulartrajectory} & human trajectory prediction \newline speed up & conditioned on observed scene & DDIM & ETH et al. & 2024 CVPR & \href{https://github.com/inhwanbae/SingularTrajectory}{SingularTrajectory} \\
\hline
\rowcolor{LightOrange}
IDM \cite{liu2024intention} & human trajectory prediction \newline speed up & conditioned on observed trajectories, endpoint & DDPM & SDD et al. & 2024 arxiv & —— \\
\hline
\rowcolor{LightOrange}
LADM \cite{lv2024learning} & human trajectory prediction \newline speed up & conditioned on coarse future trajectory & VAE \newline DDPM & ETH et al. & 2024 TIM & —— \\
\hline
\rowcolor{LightOrange}
BCDiff \cite{li2024bcdiff} & human trajectory prediction \newline instantaneous trajectory prediction & conditioned on gate & DDPM & SDD et al. & 2024 NIPS & —— \\
\hline
\rowcolor{LightOrange}
MotionDiffuser \cite{jiang2023motiondiffuser} & multi-agent prediction & conditioned on observed scene, constraints; \newline classifier guidance & LDM & WOMD & 2023 CVPR & —— \\
\hline
\rowcolor{LightOrange}
SceneDiffusion \cite{balasubramanian2023scenediffusion} & multi-agent prediction & conditioned on observed scene, interval time; \newline unconditional & LDM & Argoverse & 2023 ITSC & —— \\
\hline
\rowcolor{LightOrange}
Equidiff \cite{chen2023equidiff} & vehicle trajectory prediction & conditioned on observed trajectories, interactions & DDPM & NGSIM & 2023 ITSC & —— \\
\hline
\rowcolor{LightOrange}
Yao et al. \cite{yao2023graph} & vehicle trajectory prediction & conditioned on observed trajectories, map & DDPM & Argoverse2 & 2023 CSIS-IAC & —— \\
\hline

\end{tabular}
\end{table*}

\begin{table*}[!th]
\setlength\tabcolsep{2pt}
\centering
\renewcommand{\arraystretch}{1.5}
\begin{tabular}{|p{2.2cm}|p{3.5cm}|p{3.5cm}|p{1.5cm}|p{2.4cm}| p{1.8cm}|p{2.2cm}|}
\hline

\rowcolor{LightYellow}
Diffuser \cite{janner2022planning} & behavior planning & unconditional; \newline classifier guidance & ADM & D4RL & 2022 ICML & \href{https://github.com/jannerm/diffuser}{diffuser} \\
\hline
\rowcolor{LightYellow}
Decision Diffuser \cite{ajay2022conditional} & decision making \newline behavior planning & conditioned on rewards, constraints, skills; \newline classifier-free guidance & ADM & D4RL & 2023 ICLR & —— \\
\hline
\rowcolor{LightYellow}
MPD \cite{carvalho2023motion} & motion planning & unconditional; \newline classifier guidance & DDPM & PointMass2D & 2023 IROS & \href{https://github.com/jacarvalho/mpd-public}{mpd} \\
\hline
\rowcolor{LightYellow}
Diffusion-ES \cite{yang2024diffusion} & motion planning & unconditional & truncated DDPM & nuPlan & 2024 CVPR & \href{https://github.com/bhyang/diffusion-es}{diffusion-es} \\
\hline
\rowcolor{LightYellow}
Drive-WM \cite{wang2024driving} & motion planning \newline multiview video generation & conditioned on adjacent views & VLDM & nuScenes & 2024 CVPR & \href{https://github.com/BraveGroup/Drive-WM}{Drive-WM} \\
\hline
\rowcolor{LightYellow}
GenAD \cite{yang2024generalized} & motion planning \newline multiview video generation & conditioned on past frame, text & VLDM & WOMD et al. & 2024 CVPR & \href{https://github.com/OpenDriveLab/DriveAGI}{DriveAGI} \\
\hline

\rowcolor{LightGreen}
CTG \cite{zhong2023guided} & vehicle trajectory generation & conditioned on observed scene; \newline STL-based guidance & ADM & nuScenes & 2023 ICRA & \href{https://github.com/NVlabs/CTG}{CTG} \\
\hline
\rowcolor{LightGreen}
CTG++ \cite{zhong2023language} & multi-agent trajectory generation & conditioned on observed scene; \newline language-based guidance & ADM & nuScenes & 2023 CoRL & \href{https://github.com/NVlabs/CTG}{CTG++} \\
\hline
\rowcolor{LightGreen}
Dragtraffic \cite{wang2024dragtraffic} & multi-agent trajectory generation & conditioned on initial scene, text & LED & WOMD & 2024 IROS & \href{https://github.com/chantsss/Dragtraffic}{Dragtraffic} \\
\hline
\rowcolor{LightGreen}
DJINN \cite{niedoba2024diffusion} & multi-agent trajectory generation & conditioned on arbitrary state; \newline classifier-free guidance; \newline behavior classes guidance & EDM & Argoverse \newline INTERACTION & 2024 NIPS & —— \\
\hline
\rowcolor{LightGreen}
Pronovost et al. \cite{pronovost2023scenario} & multi-agent trajectory generation & conditioned on map, tokens & EDM \newline LDM & Argoverse2 & 2023 NIPS & —— \\
\hline
\rowcolor{LightGreen}
Rempe et al. \cite{rempe2023trace} & human trajectory generation & conditioned on observed scene; \newline classifier-free guidance & ADM & ETH et al. \newline nuScenes & 2023 CVPR & \href{https://github.com/nv-tlabs/trace}{trace} \href{https://github.com/nv-tlabs/pacer}{pacer} \\
\hline

\rowcolor{LightCyan}
FDM \cite{harvey2022flexible} & image-based driving scenario generation & conditioned on previously sampled frames & FDM & Carla & 2022 NIPS & —— \\
\hline
\rowcolor{LightCyan}
GAIA-1 \cite{hu2023gaia} & image-based driving scenario generation & conditioned on past image, text, action tokens; \newline classifier-free guidance & VDM \newline FDM & real-world dataset & 2023 arxiv & —— \\
\hline
\rowcolor{LightCyan}
DriveDreamer \cite{wang2023drivedreamer} & image-based driving scenario generation & conditioned on image, road structure, text & LDM \newline VLDM & nuScenes & 2023 arxiv & \href{https://github.com/JeffWang987/DriveDreamer}{DriveDreamer} \\
\hline
\rowcolor{LightCyan}
DriveDreamer-2 \cite{zhao2024drivedreamer} & image-based driving scenario generation & conditioned on structured info by LLMs, text & EDM & nuScenes & 2024 arxiv & \href{https://github.com/f1yfisher/DriveDreamer2}{DriveDreamer2} \\
\hline
\rowcolor{LightCyan}
Panacea \cite{wen2024panacea} & image-based driving scenario generation & conditioned on image, text, BEV sequence & LDM \newline DDIM & nuScenes & 2024 CVPR & \href{https://github.com/wenyuqing/panacea}{panacea} \\
\hline
\rowcolor{LightCyan}
DrivingDiffusion \cite{li2023drivingdiffusion} & image-based driving scenario generation & conditioned on key-frame, optical flow prior, text, 3D layout & VDM \newline LDM & nuScenes & 2023 arxiv & \href{https://github.com/shalfun/DrivingDiffusion}{DrivingDiffusion} \\
\hline
\rowcolor{LightCyan}
WoVoGen \cite{lu2023wovogen} & image-based driving scenario generation & conditioned on past world volumes, actions, text, 2D image feature & LDM & nuScenes & 2023 arxiv & \href{https://github.com/fudan-zvg/WoVoGen}{WoVoGen} \\
\hline
\rowcolor{LightCyan}
LiDMs \cite{ran2024towards} & point cloud-based driving scenario generation & unconditional; \newline conditioned on arbitrary data & LDM & nuScenes \newline KITTI-360 & 2024 CVPR & \href{https://github.com/hancyran/LiDAR-Diffusion}{LiDAR-Diffusion} \\
\hline
\rowcolor{LightCyan}
Copilot4D \cite{zhang2023learning} & point cloud-based driving scenario generation & conditioned on past observations, actions; \newline classifier-free guidance & D3PM \newline ADM & nuScenes et al. & 2024 ICLR & —— \\
\hline

\rowcolor{LightBlue}
KSTDiff \cite{zhou2023towards} & traffic flow generation & conditioned on urban knowledge graph, region feature, volume estimator & CARD & real-world dataset & 2023 SIGSPATIAL & \href{https://github.com/tsinghua-fib-lab/KSTDiff-Urban-flow-generation}{KSTDiff} \\
\hline
\rowcolor{LightBlue}
DiffTraj \cite{zhu2023difftraj} & GPS trajectory generation & conditioned on trip region, departure time; \newline classifier-free guidance & DDIM, ADM & real-world dataset & 2023 NIPS & \href{https://github.com/Yasoz/DiffTraj}{DiffTraj} \\
\hline
\rowcolor{LightBlue}
Diff-RNTraj \cite{wei2024diff} & GPS trajectory generation & conditioned on road network & DDPM & real-world dataset & 2024 arxiv & —— \\
\hline
\rowcolor{LightBlue}
ChatTraffic \cite{zhang2023chattraffc} & traffic flow generation & conditioned on text & LDM & text-traffic pairs dataset & 2024 arxiv & \href{https://github.com/ChyaZhang/ChatTraffic}{ChatTraffic} \\
\hline
\rowcolor{LightBlue}
Rong et al. \cite{rong2023complexity} & origin-destination flow generation & conditioned on node feature, edge feature & DDPM, ADM & real-world dataset & 2023 arxiv & —— \\
\hline

\end{tabular}
\end{table*}

\begin{table*}[!th]
\setlength\tabcolsep{2pt}
\centering
\renewcommand{\arraystretch}{1.5}
\begin{tabular}{|p{2.2cm}|p{3.5cm}|p{3.5cm}|p{1.5cm}|p{2.4cm}| p{1.8cm}|p{2.2cm}|}

\hline
\rowcolor{LightBlue2}
DiffSTG \cite{wen2023diffstg} & traffic flow forecasting & conditioned on past graph signals, graph structure & NCSN & PEMS et al. & 2023 GIS & \href{https://github.com/wenhaomin/DiffSTG}{DiffSTG} \\
\hline
\rowcolor{LightBlue2}
SpecSTG \cite{lin2024specstg} & traffic flow forecasting \newline traffic speed forecasting & conditioned on past graph signals feature, adjacency matrix & DDPM & PEMS et al. & 2024 arxiv & \href{https://anonymous.4open.science/r/SpecSTG/README.md}{SpecSTG} \\
\hline
\rowcolor{LightBlue2}
DiffUFlow \cite{zheng2023diffuflow} & traffic flow forecasting & conditioned on pass feature map, coarse-grained flow map, semantic features & DDPM & real-world dataset & 2023 CIKM & —— \\
\hline
\rowcolor{LightBlue2}
Xu et al. \cite{xu2023diffusion} & traffic flow forecasting & unconditional & DDPM & real-world dataset & 2023 ICASSP & —— \\
\hline
\rowcolor{LightBlue2}
ST-SSPD \cite{lablack2023long} & traffic flow forecasting & conditioned on past data points, temporal encoding, node identifier & DDPM & METR-LA et al. & 2023 MobiArch & —— \\
\hline
\rowcolor{LightBlue2}
Difforecast \cite{chi2023difforecast} & traffic flow forecasting \newline image generation & conditioned on past S-T image & DDPM & real-world dataset & 2023 BigData & —— \\
\hline

\rowcolor{LightIndigo}
Lin et al. \cite{lin2023origin} & origin-destination travel time estimation & conditioned on origin, destination, departure time & DDPM & real-world dataset & 2023 MOD & —— \\
\hline

\rowcolor{LightMagenta}
DiffTAD \cite{li2024difftad} & trajectory anomaly detection & unconditional & DDIM & NGSIM & 2024 KBS & —— \\
\hline
\rowcolor{LightMagenta}
VAD \cite{yan2023feature} & video anomaly detection & unconditional; \newline conditioned on original features & LDM, DDIM & CUHK Avenue et al. & 2023 ICCV & —— \\
\hline
\rowcolor{DirtyWhite}
AdVersa-SD \cite{fang2024abductive} & accident video understanding \newline accident preventing & conditioned on text, bounding boxes & LDM & MM-AU & 2024 CVPR & \href{https://github.com/jeffreychou777/LOTVS-MM-AU}{MM-AU} \\
\hline

\end{tabular}
\end{table*}
\end{document}